\documentclass[ijoc,nonblindrev]{informs3-neutral}
\usepackage[]{subfigure}
\usepackage{float}
\usepackage{pdflscape}

\DeclareMathOperator{\sign}{sgn}
\SingleSpacedXI

\usepackage{natbib}
 \bibpunct[, ]{(}{)}{,}{a}{}{,}%
 \def\bibfont{\small}%
 \def\BIBand{and}%

\TheoremsNumberedThrough 
\ECRepeatTheorems
\EquationsNumberedThrough 

\usepackage{float}
\usepackage{amsmath}
\usepackage{multirow}
\usepackage{microtype}
\usepackage{algorithm}
\usepackage{algorithmic}
\usepackage{commath}
\usepackage{mathtools}
\usepackage{enumitem}
\newcommand{\cO}{{\mathcal{O}}}

\begin{document}
%%%%%%%%%%%%%%%%

%\setcounter{page}{1}
%\DOI{10.1287/xxxx.0000.0000}%

\RUNAUTHOR{Herszterg, Poggi and Vidal}

\RUNTITLE{Two-Dimensional Phase Unwrapping via Balanced Spanning Forests}

% \TITLE{Bundling Information Goods of Decreasing Value}
\TITLE{Two-Dimensional Phase Unwrapping via Balanced Spanning Forests}

\ARTICLEAUTHORS{%
\AUTHOR{Ian Herszterg}
\AFF{School of Industrial and Systems Engineering, Georgia Institute of Technology, USA, \EMAIL{iherszterg@gatech.edu}} %, \URL{}}
\AUTHOR{Marcus Poggi}
\AFF{Departamento de Inform\'{a}tica, Pontif\'{i}cia Universidade Cat\'{o}lica do Rio de Janeiro, Brazil, \EMAIL{poggi@inf.puc-rio.br}} %, \URL{}}% Enter all authors
\AUTHOR{Thibaut Vidal}
\AFF{Departamento de Inform\'{a}tica, Pontif\'{i}cia Universidade Cat\'{o}lica do Rio de Janeiro, Brazil, \EMAIL{vidalt@inf.puc-rio.br}} %, \URL{}}
} % end of the block

\ABSTRACT{%
Phase unwrapping is the process of recovering a continuous phase signal from an original signal wrapped in the ($-\pi$,$\pi$] interval. It is a critical step of coherent signal processing, with applications such as synthetic aperture radar, acoustic imaging, magnetic resonance, X-ray crystallography, and seismic processing. In the field of computational optics, this problem is classically treated as a norm-minimization problem, in which one seeks to minimize the differences between the gradients of the original wrapped signal and those of the continuous unwrapped signal. When the L$^{0}$--norm is considered, the number of differences should be minimized, leading to a difficult combinatorial optimization problem.
We propose an approximate model for the L$^{0}$--norm phase unwrapping problem in 2D, in which the singularities of the wrapped phase image are associated with a graph where the vertices have $-1$ or $+1$ polarities. The objective is to find a minimum-cost balanced spanning forest where the sum of the polarities is equal to zero in each tree. We introduce a set of primal and dual heuristics, a branch-and-cut algorithm, and a hybrid metaheuristic to efficiently find exact or heuristic solutions. These approaches move us one step closer to optimal solutions for 2D L$^{0}$--norm phase unwrapping; such solutions were previously viewed, in the signal processing literature, as highly desirable but not achievable.
}

\KEYWORDS{phase unwrapping, signal processing, combinatorial optimization, graphs, mathematical programming, branch-and-cut, metaheuristics} %\HISTORY{}

\maketitle
%%%%%%%%%%%%%%%%%%%%%%%%%%%%%%%%%%%%%%%%%%%%%%%%%%%%%%%%%%%%%%%%%%%%%%

\vspace*{-0.5cm}
\section{Introduction}

The development and application of techniques for coherent signal processing have greatly increased in recent years. Synthetic aperture radar \citep{curlander1991synthetic}, magnetic resonance imaging \citep{glover1991three}, optical interferometry \citep{Pandit1994}, and X-ray crystallography \citep{guizar2011phase} are just a few examples of applications in which coherent processing is required. In the above-mentioned applications, the acquisition system measures a phase wrapped in the $(-\pi,\pi]$ domain due to trigonometric operators. \emph{Phase unwrapping} is the problem of recovering a continuous signal, the so-called absolute phase, from the wrapped phase data.
This problem has been extensively studied in the literature. The book of \cite{ghiglia1998two} and the article of \cite{zebker1998phase} give comprehensive reviews of seminal algorithms for this problem. The topic still remains very active today, because the inconsistencies caused by noise, under-sampled signals, and other natural discontinuities pose a significant challenge for state-of-the-art algorithms.

A classical objective for phase unwrapping is to minimize the norm of the difference between the gradients of the original wrapped signal and those of the continuous unwrapped signal \citep{ghiglia1996minimum}. In the particular case of the \mbox{L$^{0}$--norm}, the \emph{number} of differences should be minimized, leading to a difficult combinatorial optimization problem, closely related to a geometric Steiner problem with additional constraints \citep{chen2000network,chen2001two}. Early developments led to ad-hoc techniques and constructive algorithms \citep{goldstein1988satellite, sawaf2006finding}. Graph theory techniques have also been used to provide good approximate solutions in polynomial time. In particular, \cite{buckland1995unwrapping} relies on a minimum-cost matching algorithm, while \cite{chen2001two} investigate a network flow formulation.
Overall, the connection between phase unwrapping and operations research is very recent.

In this work, we formulate the L$^{0}$-norm 2D phase unwrapping problem as a minimum cost spanning forest problem with additional balance constraints. Although $\mathcal{NP}$-hard, this formulation is more tractable than the original Steiner forest problem. Furthermore, in practice, it produces an unwrapped solution of high quality that respects the natural contours of the image (e.g., fractures in seismic data or cliffs in synthetic aperture radar data).
To solve this formulation, we propose a branch-and-cut algorithm, using dual ascent in a pre-processing phase to reduce the number of variables via reduced-cost fixing, and a hybrid metaheuristic based on iterated local search with integer optimization over a set partitioning formulation. The key contributions of this article are the following:

1) We present a new formulation of the L$^{0}$--norm 2D phase unwrapping problem as a minimum cost spanning forest problem. This formulation aims to connect \emph{residue} points (far less numerous than the number of pixels) and seeks to better respect the natural discontinuities of the image.

2) We propose efficient exact and heuristic approaches. As our computational experiments show, our branch-and-cut algorithm can solve medium scale problems with up to $128$ residues in less than one hour on a modern computer, but its computational effort becomes impracticable for larger instances.
The metaheuristic returns high quality solutions for all our test instances with up to $1024$ residues, also retrieving most of the known optimal solutions in a more controlled CPU time.

3) Finally, this is the first application of metaheuristics and advanced mathematical programming techniques for the L$^{0}$--norm phase unwrapping problem. Promising lines of research stand wide open at the interface of these two fields.

\section{Phase Unwrapping Problem}

\begin{figure}[htbp]
    \centering
    \subfigure[]{\includegraphics[scale=0.3]{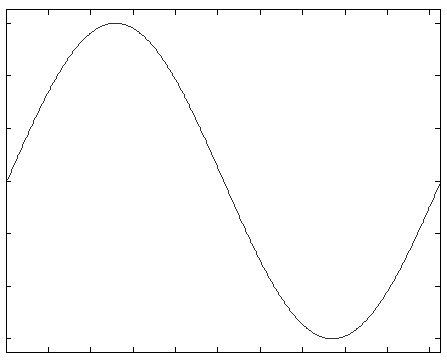}} \hspace*{0.7cm}
    \subfigure[]{\includegraphics[scale=0.3]{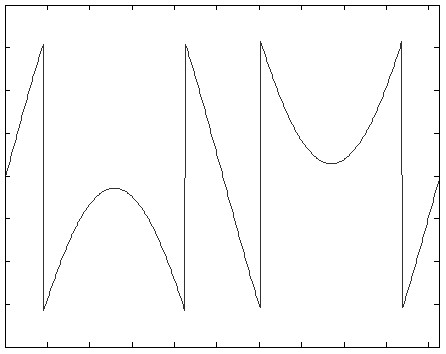}}
    \caption{Wrapping effect on a 1D continuous phase signal: (a) Continuous signal (b) Wrapped signal.}\label{fig:wrap}
\end{figure}

A large class of signal acquisition techniques, e.g., based on interferometry, produce a phase signal wrapped in the $(-\pi,\pi]$ domain by the arctangent operator.
In mathematical terms, the wrapped phase associated to a signal $\phi$(t) can be expressed as
    \begin{equation}
        W(\phi(t)) = \phi(t) + 2\pi k_{\phi}(t), \label{unwrap-eq}
    \end{equation} 
where the function $k_{\phi}(t) = \lfloor{\frac{\pi - \phi(t)}{2 \pi}} \rfloor$ wraps the phase values $\phi$(t) around $(-\pi,\pi]$, \mbox{$W$ is} the wrapping operator, and $\psi(t) = W(\phi(t))$ is the wrapped output. Figure \ref{fig:wrap} shows a continuous 1D signal and its wrapped counterpart. \vspace*{-0.22cm}

The artificial $2\pi$ \emph{jumps} of the wrapped signal $\psi$(t) must be detected and eliminated in order to reconstruct the original continuous signal $\phi$(t). This process is called \emph{phase unwrapping}. As demonstrated by \cite{itoh1982analysis}, an unambiguous phase unwrapping is possible if and only if the difference between any two adjacent samples in the continuous phase does not exceed $\pi$ (this is known as  Itoh's condition).

Consider a sequence $ \phi(1),\dots,\phi(n)$ of phase values, and define the linear differences between adjacent samples as $\Delta{\phi_{n}} = \phi(n) - \phi(n-1)$. Hence, 
\begin{equation}
\sum_{n=1}^{m} \Delta\phi_{n} = \phi(m) - \phi(1).  \label{myeq1}
\end{equation}

\noindent
Let $\Delta \psi_n =  W(\phi(n)) - W(\phi(n-1))$ be the difference between adjacent wrapped samples:
\begin{equation}
\begin{split}
  \Delta \psi_n & = [ \phi(n) + 2\pi k_\phi(n) ] - [ \phi(n-1) + 2\pi k_\phi(n-1) ] \\
    & =  \Delta_{\phi_{n}}  - 2\pi [k_\phi(n) - k_\phi(n-1)].
     \end{split}
\end{equation} 
Now, if we apply the wrapping operator $W$ over $\Delta \psi_n$, we obtain the wrapped difference between wrapped phase samples: 
\begin{equation}
      W(\Delta \psi_n)  = \Delta_{\phi_{n}} - 2\pi[k_\phi(n) - k_\phi(n-1)] - 2\pi k',
\end{equation}
where $k'$ is the proper 2$\pi$ multiple that brings the right-hand side of the equation into the $(-\pi,\pi]$ interval, ensuring that it does not violate Itoh's condition. Since $\Delta_{\phi_{n}}$ is at most equal to $\pi$ in absolute value, $k'$ must be equal to $k_\phi(n) - k_\phi(n-1)$ to keep both sides of the equation in the appropriate domain. Hence, we have
\begin{equation}
    W(\Delta \psi_n) = \Delta_{\phi_{n}},  \label{myeq2}
\end{equation} 
and substituting this expression in Equation (\ref{myeq1}) leads to
\begin{equation}
    \phi_{m} = \sum_{n=i}^{m} W(\Delta \psi_n) +\phi_{i}.  \label{myeq3}
\end{equation}

\begin{figure}[htbp]
    \hspace*{-0.5cm}
    \includegraphics[scale=0.37]{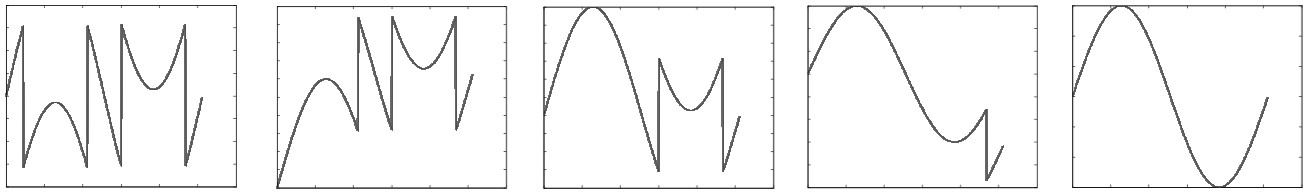}
    \caption{Unwrapping by Itoh's method for 1D phase unwrapping: the process recovers the original continuous signal from the wrapped phase samples.}
    \label{fig:unwrapping_p}
\end{figure}

As shown by \cite{itoh1982analysis}, Equation (\ref{myeq3}) can be used to derive the continuous phase $\phi_m$ from the wrapped phase along any path connecting a sample $i$ to another sample $m$. This is done by fixing the continuous phase value for an origin sample $\psi(1)$ and then adding, iteratively, each wrapped phase difference to produce the continuous phase value at the next increment. If the wrapped phase difference between consecutive samples violates the $(-\pi,\pi]$ interval, then an additional $\pm2\pi$ increment is added. These steps are executed until all the samples are evaluated. The term \emph{integration path} is used to refer to the sequence of phase values considered in this process, and Figure~\ref{fig:unwrapping_p} illustrates the procedure.

%%%%%%%%%%%%%%%%%%%%%%%%%%%%%%%%%%%%%%%%%%%%%%%%%%%%%%%%%%%%%%%%%%%%%%%%
%%%%%%%%%%%%%%%%%%%%%%%%%%%%%%%%%%%%%%%%%%%%%%%%%

\begin{figure}[htbp]
    \centering
    \subfigure[]{\includegraphics[scale=0.30]{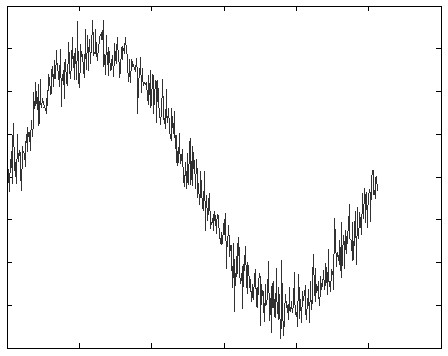}} \hspace*{0.5cm}
    \subfigure[]{\includegraphics[scale=0.30]{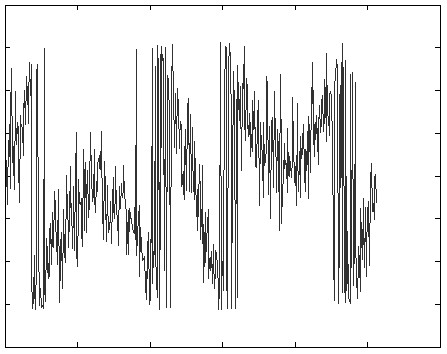}} \hspace*{0.5cm}
    \subfigure[]{\includegraphics[scale=0.30]{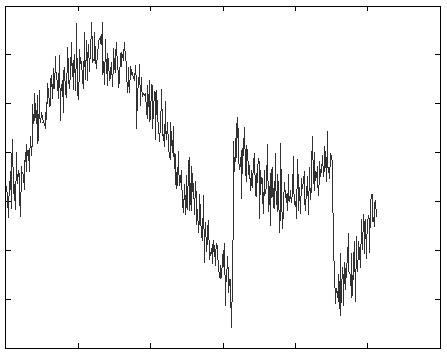}}
    \caption{Unwrapping process for noisy data: (a) Continuous phase signal with noise. (b) Wrapped noisy phase signal. (c) Unwrapping obtained by Itoh's algorithm.}
    \label{fig:noisy}
\end{figure}

Phase unwrapping problems often come from complex applications with rich geometries and signal acquisition methods that are highly susceptible to noise. Abrupt changes in the phase values can have natural causes, such as fractures or cliffs.
In this context, it is unrealistic to assume that Itoh's condition will be preserved between all adjacent samples.
Since the unwrapping process iteratively computes the difference between adjacent samples of the wrapped phase signal, any \emph{fake wrap} caused by a singularity in the data will generate an undesirable 2$\pi$ increment, which will be propagated to all subsequent samples.
Without any additional information, even the most sophisticated phase unwrapping methods will be misled. Figure~\ref{fig:noisy} shows the unwrapping process for noisy and under-sampled data, demonstrating that, in this case, Itoh's algorithm fails to reconstruct the original continuous signal.

While there is no way to overcome these issues in one dimension, an $N$-dimensional environment offers the possibility of the appropriate selection of integration paths.

%%%%%%%%%%%%%%%%%%%%%%%
%%%%%%%%%%%%%%%%%%%%%%%%%%%%%%%%%%%%%%%%%%%%%%%%%%%%%%%%%%%%%%%%%%%%%%

\noindent
\textbf{Phase unwrapping in two dimensions.}
An $N$-dimensional environment (for $N \geq 2$) 
offers more choices of integration paths, and this freedom may be exploited to better avoid damaged regions. When Itoh's condition is respected, any integration path covering all samples (2D pixels) produces the same unwrapped output. Otherwise, the unwrapping procedure is said to be \emph{path dependent}, and the choice of integration path is critical.

The issue of path dependency was first discussed by \cite{ghiglia1987cellular}.
The authors noted that singularities are restricted to certain regions in the wrapped phase data. 
The term ``residue'' was coined by \cite{goldstein1988satellite} as an analogy between the residues found in complex signals and the singularities of the phase unwrapping problem, and \cite{ghiglia1998two} showed that each residue can cause only a $\pm$2$\pi$ unwrapping error in the subsequent pixels of an integration path.
They proposed a procedure that identifies the locations of all the residues: by applying the integration scheme of Equation (\ref{myeq3}) around every elementary 2$\times$2--pixels loop, a residue is located (at the center of the loop) whenever the sum of the wrapped phase gradients differs from zero and becomes $\pm$2$\pi$. The sign of the sum defines the \emph{charge} of the residue: \emph{positive} ($+1$) or \emph{negative} ($-1$).

\begin{figure}[htbp]
    \centering
     \subfigure[]{\includegraphics[scale=0.38]{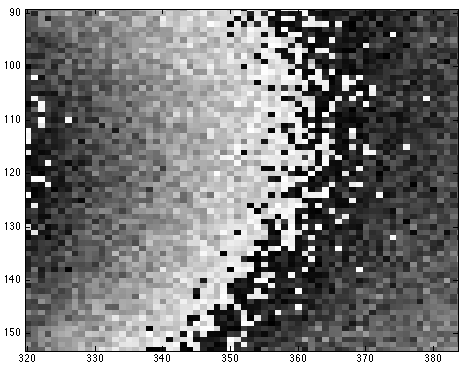}}\hspace{2em}
     \subfigure[]{\includegraphics[scale = 0.38]{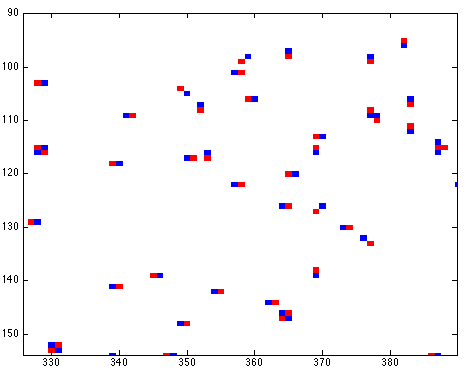}}
    \caption{Residues detected in a wrapped phase image corrupted by noise: (a) Wrapped phase image with noise. (b) Locations of the positive (blue in online article) and negative (red) residues.}
    \label{fig:residues}
\end{figure}

Figure~\ref{fig:residues} depicts the residues found in noisy wrapped phase data. An unwrapping without ambiguity (path dependent) is possible if and only if every integration path encircles none or a balanced number of residue charges.

\begin{figure}[h]
    \centering 
    \subfigure[]{\fbox{\includegraphics[height=4.9 cm, width=4.9 cm]{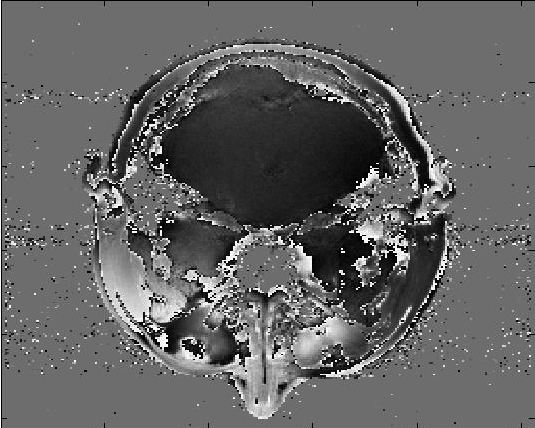}}}
    \hspace{1em}
        \subfigure[]{\fbox{\includegraphics[height=4.9 cm, width=4.9 cm]{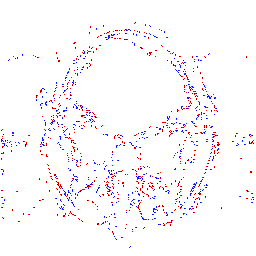}}}
                      \hspace{1em}
    \subfigure[]{\fbox{\includegraphics[height=4.9 cm, width=4.9 cm]{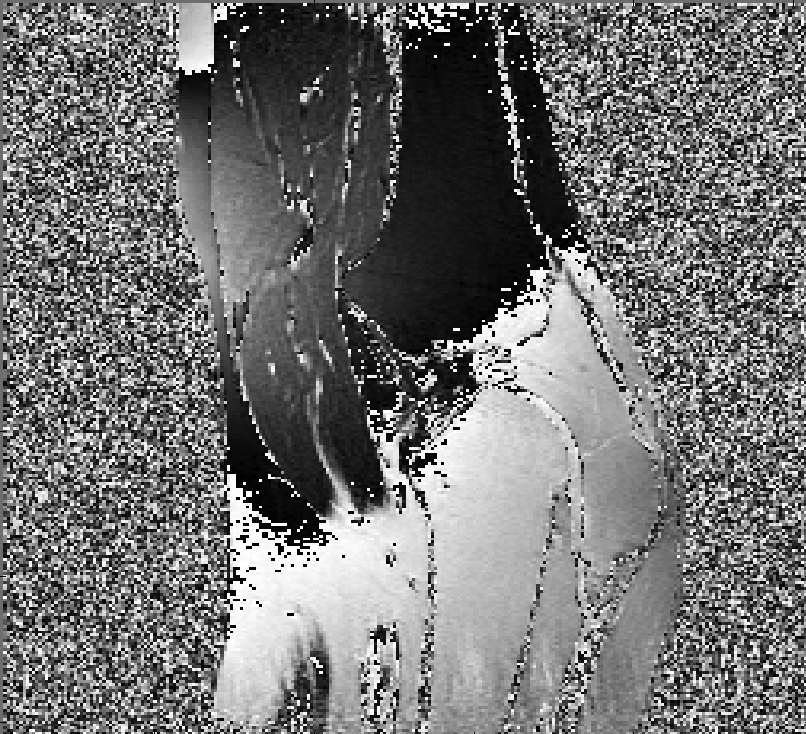}}}
        \hspace{1em}
        \subfigure[]{\fbox{\includegraphics[height=4.9 cm, width=4.9 cm]{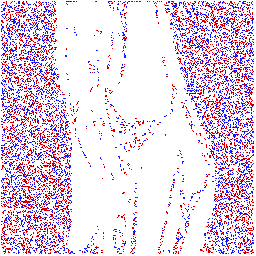}}}
    \caption{Residues detected in several wrapped phase examples from \cite{ghiglia1998two}. Many of the residues relate to contours around regions of the images. (a) Wrapped phase of a magnetic resonance image, representing a head. (b) Residues detected in (a). (c) Wrapped phase of a magnetic resonance image, representing a knee. (d) Residues detected in (c).}
    \label{fig:head_pu}
\end{figure}

While the residues of a 2D wrapped phase image indicate the existence of ambiguities, not every residue represents an error produced by noise or under-sampling. Phase discontinuities are naturally present in many phase unwrapping applications: the continuous phase values can abruptly increase or decrease in adjacent pixels. The elevations and deformities of the terrain's surface may lead to natural phase discontinuities and residues and the topology of the residues may reflect structural boundaries (see Figure~\ref{fig:head_pu} for an example). Other sources of natural discontinuities are discussed by \mbox{\cite{huntley1995characterization}}.\\

%%%%%%%%%%%%%%%%%%%%%%%%%%%%%%%%%%%%%%%%%%%%%%%%%%%%%%%%%%%%%%%%%%%%%%%%

\noindent
\textbf{Path-Following Methods.}
When Itoh's condition is not satisfied, different integration paths may lead to different unwrapped solutions because of path dependency. 
The techniques employed to resolve these ambiguities concentrate on creating artificial barriers called \emph{branch-cuts} to eliminate incorrect paths. If the branch-cuts are positioned in such a way that no integration path can encircle an unbalanced number of residues, then the path dependency is resolved. Therefore, the placement of the branch-cuts characterizes the solution, and has a visible impact on the unwrapped output.

\begin{figure}[htbp]
    \centering 
    \subfigure[]{\fbox{\includegraphics[scale=0.16]{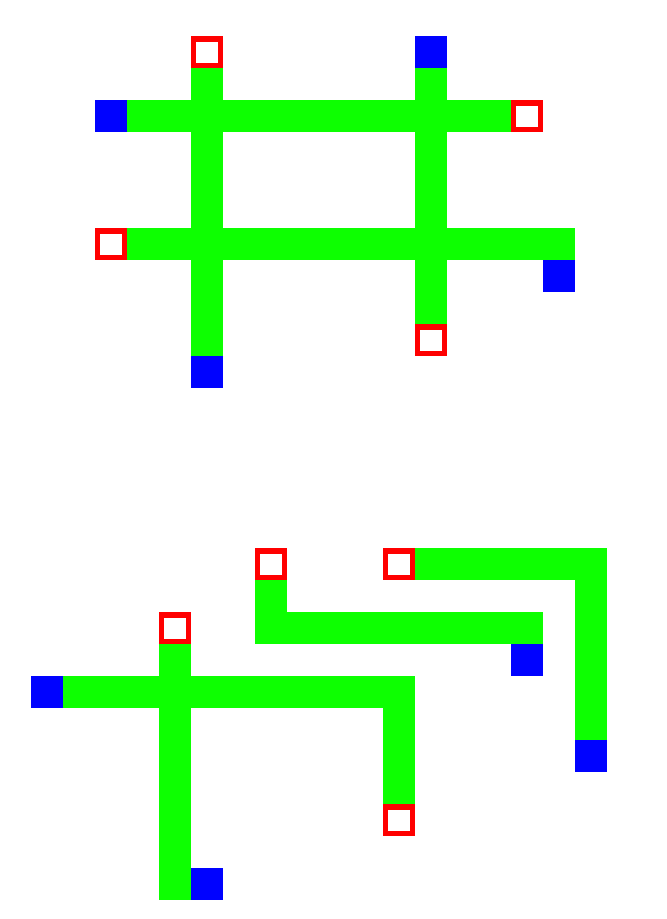}}}
    \hspace{2em}
    \subfigure[]{\fbox{\includegraphics[scale=0.16]{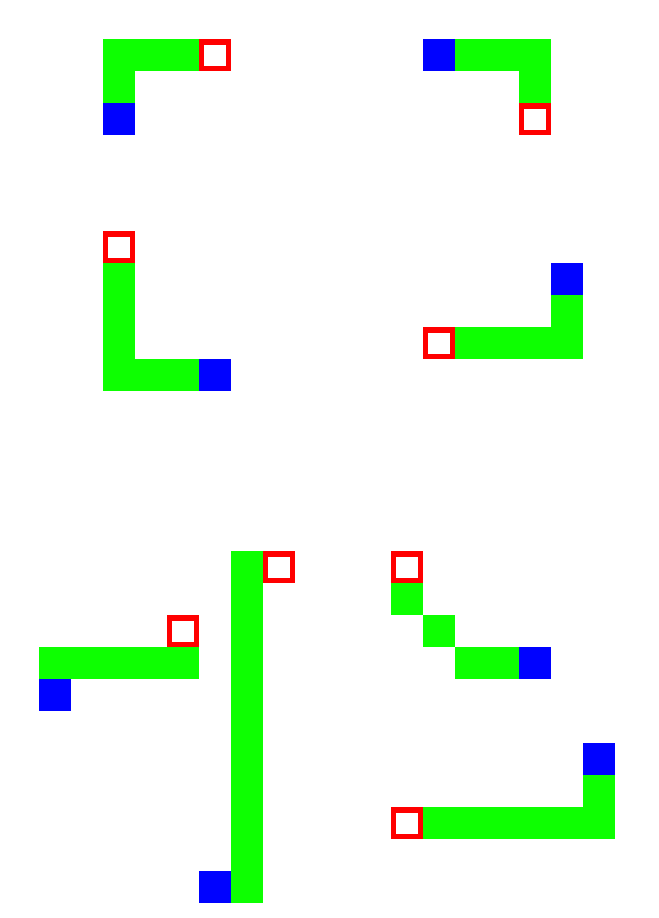}}}
    \hspace{2em}
    \subfigure[]{\fbox{\includegraphics[scale=0.16]{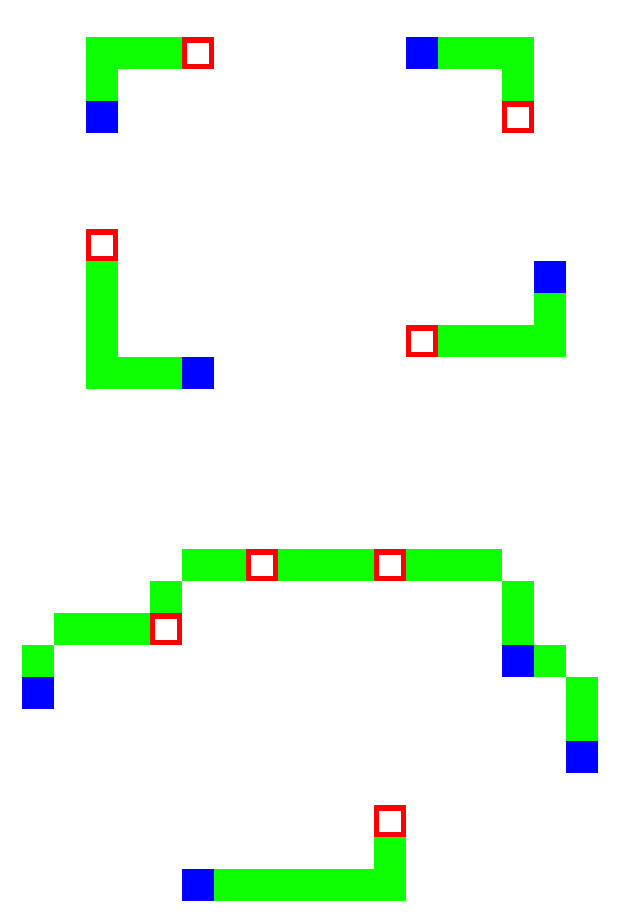}}}
    \caption{Three possible branch-cut configurations. Positive residues are represented with solid blue squares, and negative residues are represented with empty red squares. (a) First configuration, creating an isolated region. (b)--(c) Two alternative configurations with smaller overall branch-cut length.}
    \label{fig:b1}
\end{figure}

Figure~\ref{fig:b1} shows an example with eight residues. On the left, the branch-cuts represented by the green lines eliminate the path dependency. However, it is clear that they are not strategically placed, since they create an isolated region that will never be reached by an integration path. These isolated pixels would require a separate unwrapping, using an arbitrary initial phase value on a new starting sample. 

%%%%%%%%%%%%%%%%%%%%%%%%%%%%%%%%%%%%%%%%%%%%%%%%%%%%%%%%%%%%%%%%%%%%%%
%%%%%%%%%%%%%%%%%%%%%%%%%%%%%%%%%%%%%%%%%%%%%%%%%%%%%%%%%%%%%%%%%%%%%%

Finding a \emph{most-desirable} placement of the branch-cuts is a challenging task.
Figures \ref{fig:b1}(a) and \ref{fig:b1}(b) present branch-cut solutions with pairwise matchings of residues, while Figure~\ref{fig:b1}(c) illustrates a valid configuration in which several pairs of residues are connected together to achieve a smaller total length. Such a connection is possible as long as each branch-cut connects a balanced number of positive and negative residues.
Each configuration leads to a viable continuous signal after path integration.

However, some configurations are far more likely than others. In the absence of additional information such as quality maps, the most likely candidate should be the continuous signal for which the norm of the difference between the absolute and wrapped phase gradients is minimized.
Similar assumptions are used in other domains, e.g., in the phylogeny and genome comparison problems \citep{swofford1990phylogeny,miller2001comparison}, the most likely solution is likewise the one with the least number of differences according to a specific metric. 
Since only pairs of pixels from opposite sides of the branch-cuts can have a phase discontinuity of $\pm 2\pi$, minimizing the length of the branch-cuts directly relates to the minimization of the L$^{0}$--norm, where the \emph{number of differences} between wrapped and continuous phase gradients is minimized \citep{goldstein1988satellite}.

Therefore, in a general configuration, the branch-cuts can form a forest (i.e., a set of trees), such that each tree contains a balanced number of positive and negative residues.
Minimizing the total length of the trees is known to be an $\mathcal{NP}$-hard optimization problem \citep{chen2000network}, which generalizes the geometric Steiner-tree problem.
For this reason, existing algorithms do not usually seek optimal solutions but instead concentrate on restrictions of the problem that are polynomial; e.g., finding a minimum-cost matching between the positive and negative residues \citep{buckland1995unwrapping}. As shown in the next section, our contributions are an alternative simplified model, which remains $\mathcal{NP}$-hard but is more tractable, and efficient combinatorial optimization approaches to solve it.

\section{Model}
\label{section:model}

\begin{figure}[H]
    \centering
    \fbox{\includegraphics[scale=0.21]{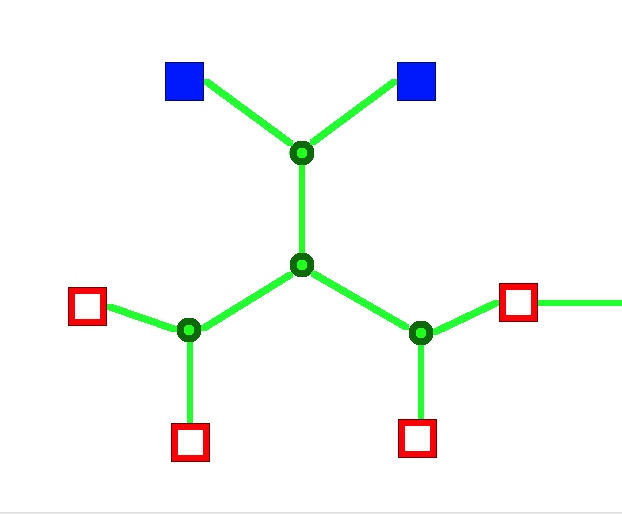}}
    \hspace{1.5em}
   \fbox{\includegraphics[scale=0.21]{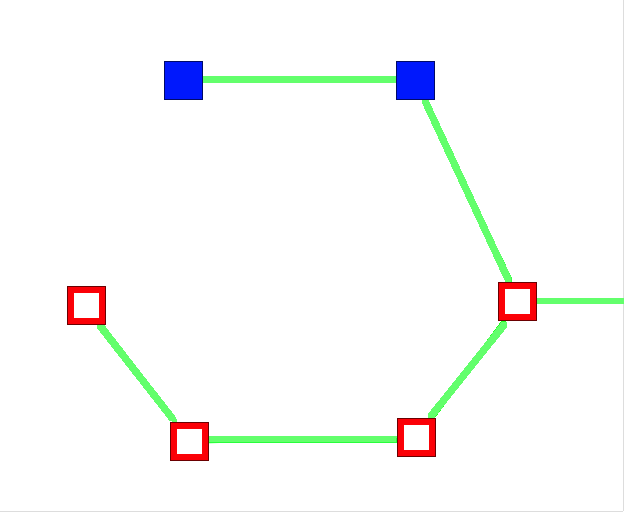}} \vspace{1.5em}
    \caption{Connecting residues with a minimum Steiner tree (left) or spanning tree (right). A Steiner tree configuration allows the use of Steiner points (green dots) to connect the residues.}
    \label{fig:steinerex}
\end{figure}

We introduce an alternative model that provides a good approximation of the problem. This model involves minimum spanning trees rather than geometric Steiner trees to connect the residues. This important difference is illustrated in Figure~\ref{fig:steinerex}.

%%%%%%%%%%%%%%%%%%%%%%%%%%%%%%%%%%%%%%%%%%%%%%%%%%%%%%%%%%%%%%%%%%%%%%%%
%%%%%%%%%%%%%%%%%%%%%%%%%%%%%%%%%%%%%%%%%%%%%%%%%%%%%%%%%%%%%%%%%%%%%%

\noindent
The motivation for this approach is fourfold:
\begin{itemize}[leftmargin=*]
\item Spanning trees may better respect the topology of the residues and the structural boundaries of the image.
\item For most practical purposes, the optimal spanning tree solution is a high-quality approximation of the Steiner solution. 
\item Branch-cuts configurations obtained via pairwise matchings \citep{buckland1995unwrapping} are special cases of spanning forests, and thus an optimal resolution of the proposed model guarantees equal or better solutions.
\item The model remains $\mathcal{NP}$-hard (see online supplement -- based on a reduction from the Steiner tree problem), but more efficient combinatorial optimization methods can be developed.
\end{itemize}
Moreover, we encourage the reader to see the \emph{Long's Peaks} data set in Section~5.2, which represents a configuration of residues from a synthetic aperture radar application. In this data set, little can be gained by allowing Steiner solutions (i.e., by creating Steiner points). The approximation related to the use of spanning trees is likely compensated by the fact that exact and heuristic methods are much more efficient on the simplified problem, leading to better solutions in less CPU time.\\

\noindent
\textbf{Problem statement.}
We now provide a formal definition of the proposed problem.
Let $G=(V,E)$ be a complete undirected graph with positive edge costs, in which each vertex $i \in V$ represents a residue with weight $w_i \in \{-1,1\}$ equal to its charge. The sum of the weight of the residues equals zero.
Each edge $\{i,j\} \in E$ represents a direct connection between residues $i$ and $j$ with distance-cost $d_{ij}$. The \emph{minimum spanning forest with balance constraints problem} (MSFBCP) aims to find a set of trees (i.e., a forest)~in~$G$, such that:
\begin{enumerate}
\item the sum of the weights of each tree is equal to zero;
\item every vertex belongs to one tree, and
\item the total cost of the trees is minimized.
\end{enumerate}
Any solution of the MSFBCP corresponds to a branch-cut configuration with the same number of branch-cuts as the number of trees.\\

\begin{figure}[htbp]
\centering
\includegraphics[scale=1.1]{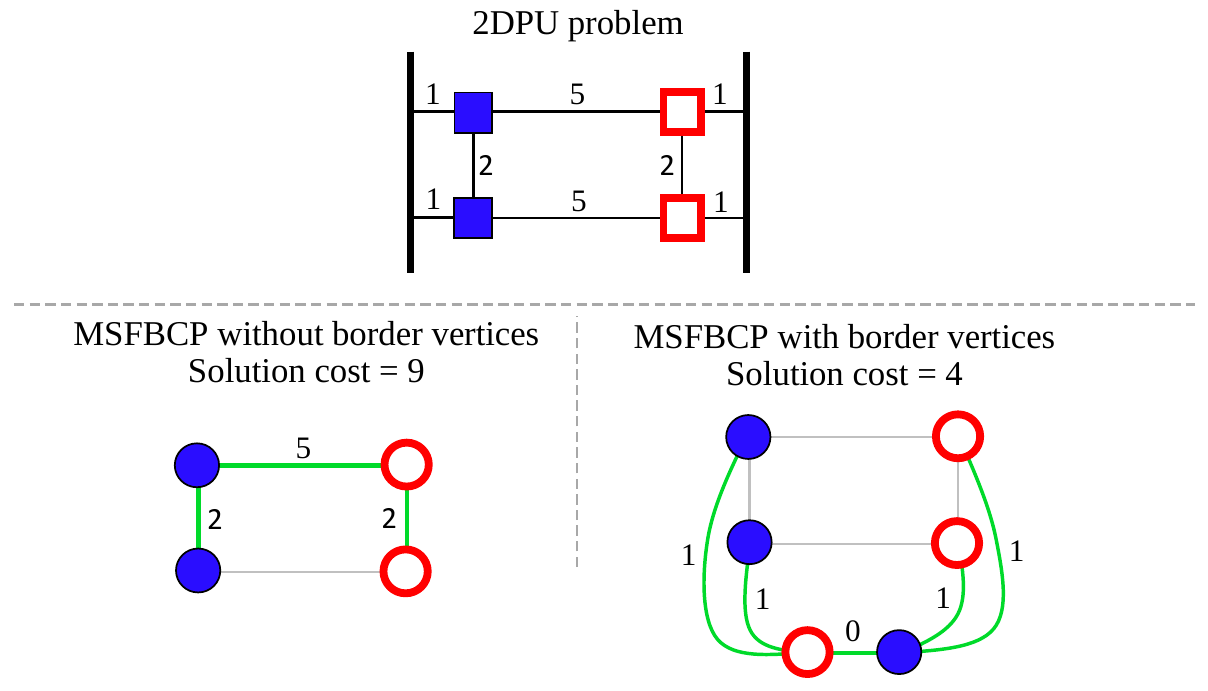}
\caption{Reformulation of the 2DPU problem into a MSFBCP, with or without border vertices}
\label{fig:2dpu_msfbc}
\end{figure}

\noindent
\textbf{Considering image borders.}
Finally, note that 2DPU applications involve finite images, and several data sets may contain an unbalanced number of positive and negative residues.
In such situations, the branch-cuts should be allowed to reach hypothetical residues beyond the borders of the image. This can be taken into account when reformulating the 2DPU into a MSFBCP, with the inclusion of a few additional vertices. Let $W$ be the sum of the residue charges in the original 2DPU problem. We add $|W|$ border vertices of charge $-\sign(W)$, as well as an additional pair of border vertices with charge $1$ and $-1$. The distance between a vertex~$i$ and a border vertex $j$ corresponds to the distance between $i$ and the closest border of the image, and the distance between two border vertices is set to zero. This transformation is illustrated in Figure \ref{fig:2dpu_msfbc}. In this example, the inclusion of the border vertices allows to connect the residues to the borders since their corresponding opposite residues may be located outside image range.

%%%%%%%%%%%%%%%%%%%%%%%%%%%%%%%%%%%%%%%%%%%%%%%%%%%%%%%%%%%%%%%%%%%%%%
%%%%%%%%%%%%%%%%%%%%%%%%%%%%%%%%%%%%%%%%%%%%%%%%%%%%%%%%%%%%%%%%%%%%%%

\section{Methods}

To solve the MSFBCP, we propose an exact mathematical programming approach as well as a hybrid metaheuristic. Roughly, the proposed exact approach can be described as follows. The MSFBCP is formulated as an integer program with an exponential number of constraints. Therefore, a {\em branch-and-cut} algorithm is used to solve the resulting model. To speed-up the resolution, we conduct a pre-processing phase where variables representing connecting arcs are removed. This is done by running a dual ascent heuristic to find feasible solutions for the dual of the integer program's linear relaxation. A dual feasible solution together with a known upper bound on the MSFBCP optimal value allows fixing, by reduced cost, variables to zero, which is equivalent to removing their associated arcs from the formulation. The branch-and-cut algorithm follows by initializing the associated linear relaxation (LP) with the cuts that were binding by the end of the dual ascent. As shown in our computational experiments, this method can solve problems with up to 128 residues to optimality. To solve larger instances, we also introduce a hybrid metaheuristic, based on iterated local search with integer optimization over a set partitioning formulation.

\subsection{Mathematical formulation}

We introduce an integer linear programming model for the MSFBCP, referred as directed cut formulation. Let $G=(V,E)$ be the graph in the MSFBCP definition. Consider now a directed graph $H=(V,A)$ where each edge $\{i,j\} \in E$ has been replaced by a pair of arcs $(i,j)$ and $(j,i)$ in $A$ with cost $d_{ij} = d_{ji}$. As is customary for Steiner tree problems \citep{claus1983nouvelle}, we rely on such a directed formulation, as it yields a better linear relaxation than its undirected counterpart (proof in online supplement).
For any subset of vertices $S \subset  V$ (cut), define the \emph{in-arcs}, the \emph{out-arcs} and the \emph{weight} of $S$ as $\delta^{-}(S) = \{(i,j) \in A: i \notin S,j \in S\}$, $\delta^{+}(S) = \{(i,j) \in A: i \in S,j \notin S\}$, and $w(S) = \sum_{i \in S} w_i$ respectively. Associating a binary variable $x_{ij}$ to each arc $(i,j) \in A$, the directed cut formulation of the MSFBCP can be expressed as follows:
\begin{align}
\min &\sum_{(i,j) \in A} d_{ij} x_{ij} \label{c02} \\
\text{s.t.}  &\sum_{ (i,j) \in \delta^+ (S) } x_{ij} \geq 1 & \forall S \subset V: w(S) > 0  \label{c32}\\
                  & \sum_{ (i,j) \in \delta^- (S) } x_{ij}  \geq 1 &\forall S \subset V: w(S) < 0   \label{c42}\\
                  &x_{ij} + x_{ji}                                 \leq 1 &  \forall (i,j) \in A: i < j  \label{c27}\\
                  &x_{ij} \in \{0, 1 \}                                     & \forall (i,j) \in A  \label{c22}
\end{align}

For every cut $S$ with a strictly positive weight, Constraints~(\ref{c32}) force one or more  \emph{out-arcs} to be included in the solution.
Similarly, for every cut $S$ with a strictly negative weight, Constraints~(\ref{c42}) force one or more \emph{in-arcs} to be included in the solution.
Finally, Constraints~(\ref{c27}) prohibits the selection of a pair of edges with opposite directions. One may observe that for each set $S$ defining a constraint~(\ref{c32}) there is an equivalent constraint~(\ref{c42}) defined by set $V \setminus S$. We use both. This allows representing the feasible set of solutions, where no unbalanced spanning tree is allowed, with sparser constraints.

\subsection{Dual ascent heuristic}

The first component of our algorithm is a dual heuristic, which aims to quickly generate an initial dual solution. This solution is subsequently used to simplify the problem by variable fixing. This strategy has been successfully used for the Steiner problem in graphs \citep{wong1984dual,de2001dual}, which shares many common features with the considered problem.
The MSFBCP directed cut formulation has constraints for all unbalanced cuts~$S \subset V$. Let~$\pi_{S}$ be the \emph{dual variables} associated to Constraints~(\ref{c32})--(\ref{c42}), and~$\lambda_{ij}$ be the dual variables associated to Constraints (\ref{c27}). The dual formulation can be expressed as:
\begin{align}
\max \hspace*{0.2cm} & \hspace*{0.8cm}  \sum\limits_{S \subset V: \ w(S) \neq 0}  \hspace*{-0.4cm} \pi_{S}  \hspace*{0.3cm}  - \sum\limits_{(i,j) \in A: \ i < j} \lambda_{ij}
\label{c0d1} \\
\text{s.t.} \hspace*{0.2cm} &\sum_{S \subset V: \  w(S) > 0, (i,j) \in \delta^{+} (S)}  \hspace*{-1.3cm} \pi_{S}  \hspace*{0.8cm} + \sum_{ S \subset V: \ w(S) < 0, (i,j) \in \delta^{-} (S)}  \hspace*{-1.3cm} \pi_{S} \hspace*{0.8cm}  \leq d_{ij} +  \lambda_{ij}
& \forall (i,j) \in A   \label{c1d1} \\
  &\pi_{S} \geq 0 &\forall S \subset V: \ w(S) \neq 0  \label{c2d1a} \\
& \lambda_{ij} \geq 0 & \forall (i,j) \in A: i < j  \label{c2d1b}
\end{align}
This problem corresponds to maximizing the sum of the dual variables associated to the unbalanced cuts. Constraints~(\ref{c1d1}) ensures primal optimality conditions, i.e. the sum of the dual variables associated with the unbalanced cuts containing a given arc $a \in A$ is, at most, equal to the cost of that edge.

A dual ascent heuristic starts with a dual feasible solution. It proceeds by increasing one dual variable at a time. Since all edge costs are nonnegative, we start with a zero value duals ($\boldsymbol\pi = 0$), which is dual feasible. Duals $\boldsymbol\lambda$ are kept zero all along our procedure. As in Wong's dual ascent algorithm, we sequentially choose cuts that correspond to connected components in graph $H_{\pi}=(V,A(\pi))$, for $\pi$ the associated dual solution.  $A(\pi)$ is the set of saturated arcs, i.e. it contains only arcs with zero reduced cost for a given dual feasible solution $\pi$.
At each iteration, a directed cut is chosen and its corresponding dual is increased until at least one of its arcs is saturated. Therefore, as the number of connected components reduces by at least one, at most $|V|-1$ iterations are performed.

We select either the connected component associated to the violated cut that contains the minimum-reduced-cost edge in its edge set, or choose randomly a connected component. Both criteria will produce feasible and maximal dual solutions, but the quality of the lower bounds depends on this choice. Section~5 presents an experimental comparison of the two criteria. Experimental results indicate that the random approach generates better bounds.

At each iteration, we detect violated cuts by a depth-first search in the graph $H_{\overline{\pi}}$, in $\cO(\abs{V}+\abs{E})$ operations. 
The complexity of the selection of the violated cuts depends on the selection criterion. For the criterion of minimum reduced cost edge, it takes $\cO(|E|)$. The random cut criterion chooses a cut by a draw in constant time. In both cases, the overall dual ascent procedure runs in $\cO(\abs{V}(|V|+|E|))$.

Finally, we apply a dual scaling approach, which consists in multiplying the dual solution by a constant factor $0 < \alpha < 1$, leading to a feasible but not maximal dual solution, and applying the dual ascent starting from the dual solution obtained. This may produce a better lower bound, since new sets of dual variables are likely to assume positive values, giving cuts that were not previously considered in the dual solution. This procedure terminates when either a better lower bound is obtained or a maximum of $It_{DS}$ trial iterations has been performed, as  in \cite{de2001dual}.

It is important to stress again that the aim of this procedure is to provide fast
lower bounds. These bounds, even if not tight, enable to fix a significant
number of arc variables to zero. The reduction obtained turns out to be
a relevant step in the final algorithm.

\subsection{Branch-and-cut}

To produce optimal solutions, we propose a branch-and-cut (B\&C) algorithm.
The method starts by using the dual heuristics to reduce the size of the problem.
Given the solution of the dual ascent procedure and an upper bound provided by the best known primal solution (obtained from the heuristic of Section \ref{meta}), we first remove the set of arcs whose reduced costs exceed the gap between the upper and lower bounds.

We then initialize the search tree by solving MSFBCP's linear programming relaxation (LP) at the root node for the reduced instance. The cuts $W$ associated with positive dual variables at the end of the dual ascent procedure, i.e. $W$ is such that $\overline{\pi}_{w} > 0, \forall \>  w \in W$, are inserted as initial constraints in the LP formulation. At each search-tree node, the LP is solved by repeatedly finding violated unbalanced directed cuts through the separation procedure described later in this section. \\

\noindent
\textbf{Exploration Strategy.} \
We choose the most fractional variable as the branching variable at each node, and explore the search tree in a depth-first manner to reduce the number of active nodes and thus the overall memory requirement. After a pruning operation, the search resumes from the nearest parent node in the tree.\\

\noindent
\textbf{Solution of the LP.} \
The exponential number of constraints requires a separation procedure to solve the relaxed MSFBCP~(Equations \ref{c02}--\ref{c27}). Consider an initial LP with a subset of the unbalanced directed cuts. For instance, the ones containing only a single vertex. Its optimal solution is a minimum spanning forest where all the vertices are connected to at least one other vertex. 
Let $\overline{x}$ be the current optimal, possibly fractional, LP solution. A violated 
directed cut is an unbalanced set $S$ for which the sum of $\overline{x}$ values for all the arcs in $\delta^{+}(S)$ (resp. $\delta^{-}(S)$ for negatively unbalanced sets $S$) is smaller than one. 

Therefore, finding a violated cut amounts to determine whether
there are source-sink pairs of vertices with opposite weights such
that the maximum flow/minimum cut is smaller than one in graph $H^+=(V,A^+)$
where $A^+$ contains only arcs $a \in A$ associated with positive~$\overline{x}_a$.
The capacities of the arcs $a \in A^+$ are set to $\overline{x}_a$.
This problem can be solved efficiently in $\cO(nm \log(n^{2}/m))$ using the algorithm of \cite{goldberg1988new}. The connected components of $H^+$ correspond to violated cuts when the set of vertices of the component is not balanced. We speed up our separation within each connected component of $H^+$ by avoiding to solve the separation problem for all pairs of vertices with opposite weights. Observe that when a minimum cut $S$ between a pair of vertices is found, any pair of vertices $(i,j)$ such that $i \in S$ and $j \notin S$ will lead to the same minimum cut. Therefore, the algorithm proceeds by recursively picking pairs at each side of the minimum cut obtained previously, until attaining a single vertex or not identifying any unbalanced cut.

\subsection{Hybrid metaheuristic}
\label{meta}

Finally, to generate good MSFBCP primal solutions for large instances, we propose a hybrid iterated local search (HILS) metaheuristic. Algorithm~\ref{ILS} gives the pseudocode.

\begin{algorithm}[ht]
\SingleSpacedXI
\linespread{1.0}\selectfont
  \caption{Hybrid Iterated Local Search}
\label{ILS}
\begin{algorithmic}[1]
\STATE \textbf{input:} A graph $G = (V,E)$;
\STATE \textbf{output:} A set of balanced spanning trees $S^*$;
\STATE $S$ $\leftarrow$ GenerateInitialSolution(G); \label{algo1}
\STATE $S^* \leftarrow S$; $It_\textsc{shak} \leftarrow 0$;
\WHILE {$It_\textsc{shak} < It_\textsc{max}$} 
\STATE $S \leftarrow$ LocalSearch($S$); \label{algo2}
\IF {$\exists  \ k \in \mathbb{N}^+ \text{ s.t. } It_\textsc{shak} = k \times  It_\textsc{sp}$}
\STATE $S \leftarrow$ SetPartitioning(); \label{algo4}
\ENDIF
\IF {$c(S) < c(S^*)$}
\STATE $S^* \leftarrow S$;
\STATE $It_\textsc{shak} \leftarrow 0$;
\ENDIF
\STATE $S \leftarrow$ Perturb($S$) or Perturb($S^*$) with equal probability;  \label{algo3}    
\ENDWHILE
\STATE \textbf{return} $S^*$;
\end{algorithmic}
\end{algorithm}

After an initial solution construction (Line \ref{algo1} of Algorithm \ref{ILS}), the method iteratively applies a local search improvement procedure  (Line \ref{algo2}), followed by a perturbation procedure to escape local minima (Line \ref{algo3}). This is applied with equal probability either to the current solution or to the best solution so far. We introduced this rule to achieve a balance between diversification and aggressive solution improvement.
An integer optimization algorithm is called to solve a set partitioning formulation at every $It_\textsc{sp}$ iterations (Line \ref{algo4}).
The algorithm terminates after $It_\textsc{max}$ consecutive iterations of local search and perturbation with no improvement of the best solution.

It is important to note that this algorithm uses an indirect solution representation. Each solution is represented as a partition $(P_1,\dots,P_k)$ of the set of vertices such that $\bigcup_i P_i = V$ and $P_i \cap P_j = \varnothing$ for $i \neq j$. Each component $P_i$ represents a tree, and its cost $c(P_i)$ can be efficiently derived by solving a minimum-cost spanning tree problem. Furthermore, an unbalanced component~$P_i$ is not considered infeasible, but instead penalized with a cost of~$\xi(P_i)$. For problem instances that represent a 2DPU applications, $\xi(P_i)$ is set to the distance to the closest border of the image. In other situations, a fixed penalty can be used.\\ 

The \textbf{initial solution} is obtained by first computing a minimum-cost spanning tree for the whole vertex set $V$ and then disconnecting any edge that is longer than a threshold~$d_\textsc{max}$.\\

The \textbf{local search} procedure, applied to the initial solution as well as to any solution created by the perturbation operator, is conducted in the search space of the subsets.
The method explores several neighborhoods, obtained via simple moves of the vertex-to-subset assignment decisions. In this context, the evaluation of a neighbor solution requires the evaluation of the minimum-cost spanning trees for the modified components. All the neighborhoods of our method involve no more than two components, and thus we evaluate a maximum of two spanning trees per move.

The exploration of the neighborhoods is exhaustive, enumerating all the component pairs $(P_i,P_j)$ in random order to test the associated moves. Any improving move is directly applied, and the local search stops when no further improving moves exist. This policy saves many move re-evaluations, since it is unnecessary to attempt a move between $P_i$ and~$P_j$ if this move has been evaluated in the past and the components have not changed. The following moves are considered:
\begin{itemize}
\item[--]
\textsc{Relocate}: The neighborhood considers all relocations of one vertex from $P_{i}$ to $P_{j}$. For a given pair $(P_i,P_j)$, this neighborhood can be evaluated in \mbox{$\cO( |V_i|( \abs{E_i}\log\abs{E_i} +  \abs{E_j}\log\abs{E_j} )) $}.

\item[--]
\textsc{C-Relocate}: The neighborhood tries to relocate any pair of close
positive or negative vertices from $P_{i}$ to $P_{j}$. It can be evaluated in
 \mbox{$\cO( \abs{V_i}^2( \abs{E_i}\log\abs{E_i} +  \abs{E_j}\log\abs{E_j} )) $}.

\item[--]
\textsc{Swap}: The neighborhood considers all swaps of a pair of vertices with the same weight between $P_{i}$ and $P_{j}$. This neighborhood can be evaluated in \mbox{$\cO(\abs{V_i}\abs{V_j}( \abs{E_i}\log\abs{E_i} +  \abs{E_j}\log\abs{E_j} ))$}.

\item[--]
\textsc{C-Swap}: Similarly to C-Relocate, this neighborhood swaps pairs of positive and negative vertices between $P_{i}$ and $P_{j}$. The C-Swap neighborhood can be evaluated in \mbox{$\smash{\cO(|V_i|^2|V_j|^2( \abs{E_i}\log\abs{E_i} +  \abs{E_j}\log\abs{E_j} ))}$}. To reduce the computational effort, the second vertex of each pair is selected from a subset of the closest vertices.

\item[--]
\textsc{Merge}: The neighborhood merges a partition pair. For a given pair $(P_i,P_j)$, this neighborhood can be evaluated in \mbox{$\cO(\abs{E_i + E_j}\log\abs{E_i + E_j})$}.

\item[--]
\textsc{Break}: The neighborhood attempts to remove one edge of the minimum spanning tree of $P_{i}$ to generate two new components. This is done for all edges of the spanning tree. This neighborhood can be evaluated in $\cO(\abs{V_i}\abs{E_i}\log\abs{E_i})$.

\item[--]
\textsc{Insert 1, Break 1}: The neighborhood first merges $P_{i}$ and $P_{j}$ and computes the minimum spanning tree of the resulting component. It then removes the longest edge in such a way that two new components are formed. This neighborhood can be evaluated in \mbox{$\cO(\abs{E_i + E_j}\log\abs{E_i + E_j})$}.
\end{itemize}

To reduce the computational effort, the neighborhoods are evaluated only between vertices located within a given distance, defined in a preprocessing stage for each vertex. Similarly, a move between two components is applied only if the shortest distance between their vertices does not exceed the limit. \\

The \textbf{perturbation procedure} is applied to escape from local minima of the previous neighborhoods. This operator is applied with equal probability to either $S$ or $S^*$. The perturbation removes $k$ edges of the trees of the current solution, creating new disjoint subsets that are randomly recombined to resume the search with $k$ subsets, where $k$ is a random integer from a uniform distribution in the interval $[0, 0.15T]$, where $T$ is the number of trees.\\ 

Finally, an integer optimization algorithm over a \textbf{set partitioning} formulation is used after each $It_\textsc{sp}$ iterations, with the aim of generating an improving solution from the components of previous high-quality solutions. This formulation can be written as:
%%%%%%%%%%%%%%%%%%%%%%%%%%%%%%%%%%%%%%%%%%%%%%%%%%%%%%%%%%%%%%%%%%%%%%%
\begin{align}
\min &\sum\limits_{p \in \mathcal{P}} c(p) x_p \\
 \text{s.t.} \hspace{2mm} &\sum_{ p \in \mathcal{P} } a_{vp}x_p  = 1 & \forall v \in V \\
 & x_{p} \in \{0, 1 \} & \forall p \in \mathcal{P} \\
 &\text{where } a_{vp} = \begin{cases} 1 & \quad \text{if } v \in p\\ 0 & \quad \text{if } v \notin p,\\ \end{cases} 
\end{align}
%%%%%%%%%%%%%%%%%%%%%%%%%%%%%%%%%%%%%%%%%%%%%%%%%%%%%%%%%%%%%%%%%%%%%%%
where the set $\mathcal{P} \subseteq 2^V$ is formed of up to $P_\textsc{size}$ components belonging to the most recent local minima of HILS. If an improved solution is found, it is used as the current solution for the~HILS.

\section{Computational experiments}

Our computational experiments were designed to
1) measure the performance of our methods on a set of benchmark instances for the MSFBCP; 
2) compare the MSFBCP solutions for 2D phase unwrapping to those of other path-following methods. 

The algorithms were implemented in C++, and Gurobi v6.0.4 was used for the resolution of the linear programs.
All experiments with HILS were conducted on an Intel i7 2.3\,GHz CPU, and the experiments with the B\&C method were conducted on an Intel Xeon E5-2650v3 2.3 GHz CPU. A single thread was used.

\subsection{Solution quality for MSFBCP}

We first created a set of MSFBCP instances with a variety of vertex configurations and problem sizes. 21 groups of instances were generated, with a number of vertices ranging from $10$ to~$1026$. For each group, called ``\textsc{PUC-N}'', five instances were generated by randomly distributing $N/2$ positive vertices and $N/2$ negative vertices in an $4N \times 4N$ Euclidean space. The cost of each edge is computed as the 2D Euclidean distance between the vertices, with double precision. As described in Section \ref{section:model}, a pair of positive and negative border vertices were added in each instance to allow possible connections to the border of the Euclidean space. All instances can be found in the online supplement of this paper, or at the following address: \url{https://w1.cirrelt.ca/~vidalt/en/research-data.html}.

\subsubsection{Hybrid metaheuristic}

We conducted preliminary experiments to calibrate the parameters of the hybrid metaheuristic, with the aim of balancing the effort dedicated to the different search components and returning results in a few minutes for medium instances. Following these experiments, we set the termination criteria to $It_\textsc{max} = 100$ and $T_\textsc{max} = 3600$\,s. The set partitioning routine is performed every $It_\textsc{sp} = It_\textsc{max}/3$ iterations, with a maximum pool size of $P_\textsc{size} = 1000$ columns and a time limit of $300$ seconds. Finally, $d_\textsc{max}$ is set to the average distance of an edge in $G$, and the maximum distance radius for every vertex $v$ is set to 25\% of the shortest distance between \emph{v} and the set of vertices~$V-\{v\}$.

The HILS algorithm was then executed 10 times on each benchmark instance with different random seeds. Table \ref{tab:meta} summarizes the results for each group of instances. Column \textbf{GAP$_\textsc{Best}$(\%)} gives the average gap between the best solution of 10 runs and the \emph{best known solution} (BKS) collected from all our experiments.
 Column \textbf{GAP$_\textsc{Avg}$(\%)} gives the average gap between the average solution over the 10 runs and the BKS.  Column \textbf{OPT} gives the number of optimal solutions found by the HILS algorithm and the total number of known optimal solutions. Finally, Column \textbf{Avg T(s)} gives the average time per run and per instance, measured in seconds. 

\begin{table}[htbp]
\centering
\renewcommand{\arraystretch}{1.2}
\setlength{\tabcolsep}{.7em}
\caption{Performance of the HILS algorithm}
\label{tab:meta}
\scalebox{0.82}{
\begin{tabular}{|c|cccc|}
\hline
Group &  GAP$_\textsc{Best}$(\%) & OPT & GAP$_\textsc{Avg}$(\%) & Avg T(s) \\ \hline
PUC-8 &  0.00 &  \textbf{5/5} & 2.49 & 0.27 \\
PUC-12 &  0.00 & \textbf{5/5} & 0.00 & 0.77 \\
PUC-16 &  0.00 & \textbf{5/5} & 0.52 & 1.60 \\
PUC-20 & 0.00 &  \textbf{5/5} & 0.21 & 3.60 \\
PUC-24 &  0.00 &  \textbf{5/5} & 0.64 & 5.46 \\
PUC-28 &  0.00 &  \textbf{5/5} & 1.06 & 10.50 \\
PUC-32 &  0.00 & \textbf{5/5} & 0.76 & 14.88 \\
PUC-36 &  0.00 &  \textbf{5/5} & 0.94 & 19.91 \\
PUC-40 &  0.00 &  \textbf{5/5} & 0.63 & 32.49 \\
PUC-44 &  0.14 &  4/5 & 1.44 & 44.50 \\
PUC-48 &  0.00 &  \textbf{5/5} & 0.73 & 48.91 \\
PUC-52 &  0.00 &  \textbf{5/5} & 1.33 & 70.03 \\
PUC-56 & 0.00 &  \textbf{5/5} & 1.35 & 82.76 \\
PUC-60 &  0.00 &  \textbf{5/5} & 1.15 & 97.13 \\
PUC-64 &0.40 &  4/5 & 3.02 & 134.45 \\
PUC-80 & 0.40 &  3/3 & 3.26 & 304.64 \\
PUC-96 &  0.34 & 2/5 & 4.81 & 650.05 \\
PUC-128 &  1.78  & 2/5 & 5.45 & 2091.65 \\
PUC-256 &  0.00  & 0/0 & 5.78 & 3600.00 \\
PUC-512 &  0.00  & 0/0 & 6.31 & 3600.00 \\
PUC-1024 & 0.00  & 0/0 & 4.85 & 3600.00 \\
\hline
\end{tabular}
}
\end{table}

The method found the optimal solution in at least one run for 80 of the 105 instances. The solutions obtained for 15 other instances were considered to be the best known primal solutions, with no guarantee of optimality. The average deviation in terms of solution quality (total length of the branch-cuts) for all the instances is 0.29\%. The detailed results are given in the online supplement to this paper.

\begin{figure}[htbp]
 \centering
\includegraphics[scale=0.28]{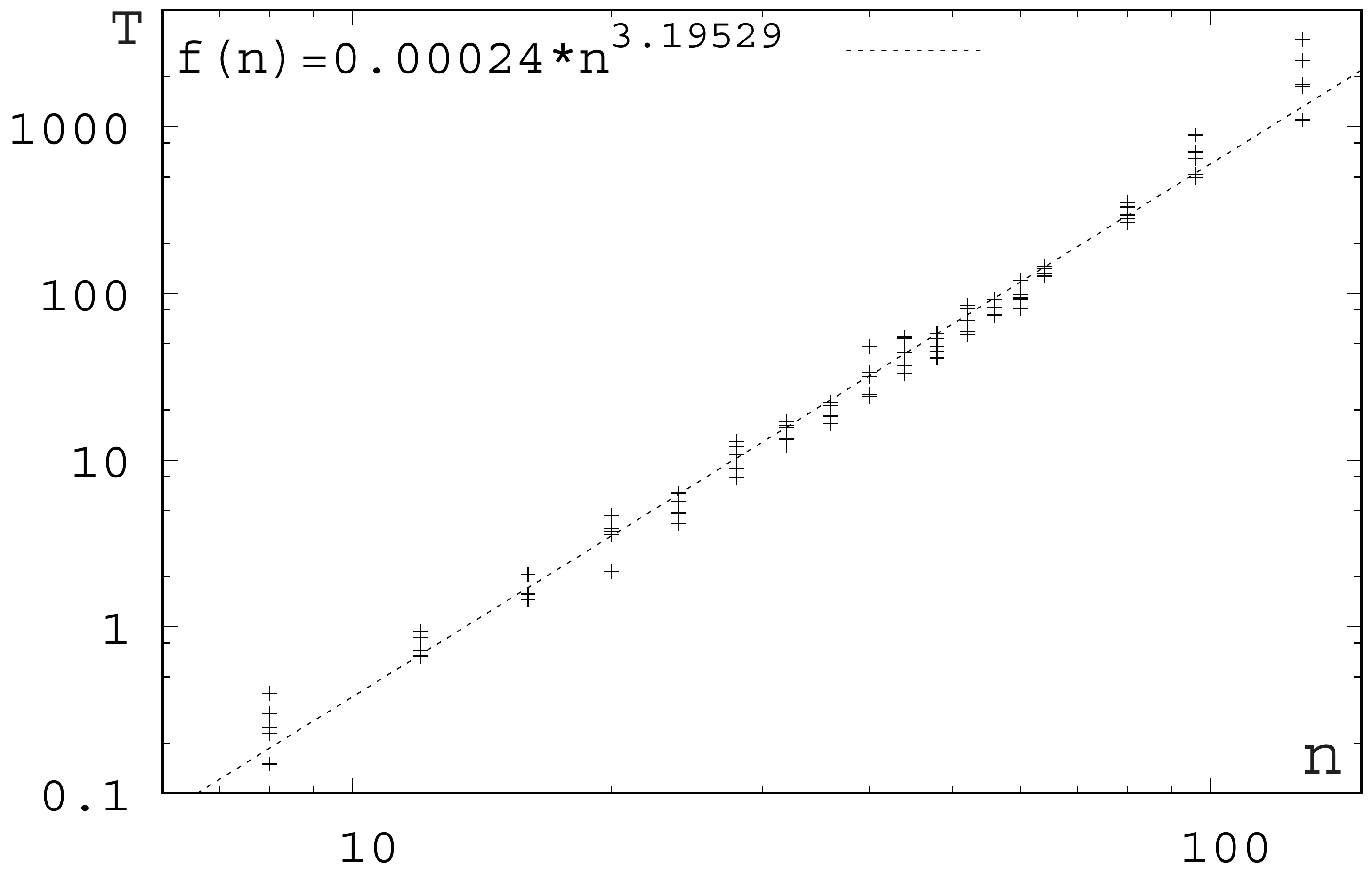}
  \caption{Growth of the CPU time of HILS as a function of problem size, log-log scale}
    \label{fig:meta_graph}
\end{figure}

The CPU time ranges from a fraction of a second for small instances to 60 minutes for the largest instances. Figure \ref{fig:meta_graph} shows the CPU time as a function of problem size for those instances in which the time limit (3600 seconds) was not reached. In this subset where the termination criterion is the number of iterations without improvement, the fitted curve (using a power law of the form $f(n) = \alpha \times n^\beta)$ suggests cubic growth of the CPU time as a function of problem size.

%%%%%%%%%%%%%%%%%%%%%%%%%%%%%%%%%%%%%%%%%%%%%%%%%%%%%%%%%%%%%%%%%%%%%%%%
%%%%%%%%%%%%%%%%%%%%%%%%%%%%%%%%%%%%%%%%%%%%%%%%%%%%%%%%%%%%%%%%%%%%%%

\subsubsection{Dual heuristics}

Table \ref{tab:dual} presents the results of the dual ascent algorithm. We compare two criteria for the selection of violated cuts: a) choosing the minimum-reduced-cost-arc in the graph (\textbf{Min-RC}), and b) randomly selecting a non-maximal dual variable and saturating at least one of its arcs (\textbf{Random}).
Columns \textbf{GAP(\%)}, \textbf{T(s)} and \textbf{R(\%)} report the average gap between the dual bounds and the (primal) BKS, the average CPU time in seconds and the percentage of problem reduction achieved by fixing arcs by reduced costs, respectively.

\begin{table}[!htbp]
\centering
\caption{Results of the dual ascent algorithm, comparing two cut-selection strategies}
\setlength{\tabcolsep}{.7em}
\renewcommand{\arraystretch}{1.2}
\label{tab:dual}
\scalebox{0.8}{
\begin{tabular}{|ccc|ccc|ccc|ccc|}
\cline{4-9}
 \multicolumn{3}{c|}{\strut} & \multicolumn{3}{c|}{\textsc{Min-RC}} & \multicolumn{3}{c|}{\textsc{Random}} \\
\hline
Group & $|V|$ & $|E|$ & GAP(\%) & Avg T(s) & R(\%) & GAP(\%) & Avg T(s) & R(\%)\\
\hline
PUC-8 & 10 & 90 & 2.10 & $<$0.01 & 92.44 & 0.00 & $<$0.01 & 94.67\\
PUC-12 & 14 & 182 & 7.34 & $<$0.01 & 90.55 & 0.00 & $<$0.01 & 99.12\\
PUC-16 & 18 & 306 & 5.02 & $<$0.01 & 93.86 & 1.00 & $<$0.01 & 97.78\\
PUC-20 & 22 & 462 & 10.83 & $<$0.01 & 83.29 & 1.20 & $<$0.01 & 97.23\\
PUC-24 & 26 & 650 & 5.52 & $<$0.01 & 93.17 & 1.28 & $<$0.01 & 97.48\\
PUC-28 & 30 & 870 & 10.13 & $<$0.01 & 83.68 & 2.05 & $<$0.01 & 97.33\\
PUC-32 & 34 & 1122 & 11.88 & $<$0.01 & 75.97 & 2.74 & $<$0.01 & 95.58\\
PUC-36 & 38 & 1406 & 21.15 & 0.03 & 56.16 & 3.87 & $<$0.01 & 94.28\\
PUC-40 & 42 & 1722 & 14.90 & 0.07 & 62.93 & 4.64 & 0.01 & 92.31\\
PUC-44 & 46 & 2070 & 16.45 & 0.04 & 55.77 & 5.33 & 0.02 & 90.14\\
PUC-48 & 50 & 2450 & 7.63 & 0.06 & 85.52 & 2.95 & 0.02 & 96.28\\
PUC-52 & 54 & 2862 & 13.22 & 0.11 & 67.58 & 4.53 & 0.03 & 92.61\\
PUC-56 & 58 & 3306 & 12.22 & 0.09 & 68.64 & 4.22 & 0.03 & 93.54\\
PUC-60 & 62 & 3782 & 13.43 & 0.14 & 62.90 & 5.25 & 0.03 & 90.08\\
PUC-64 & 66 & 4290 & 16.34 & 0.05 & 46.82 & 6.05 & 0.05 & 86.25\\
PUC-80 & 82 & 6642 & 16.89 & 0.24 & 38.77 & 6.07 & 0.09 & 84.36\\
PUC-96 & 98 & 9506 & 19.09 & 0.71 & 34.31 & 7.23 & 0.14 & 76.39\\
PUC-128 & 130 & 16770 & 14.25 & 2.36 & 25.23 & 7.12 & 0.41 & 62.17\\
PUC-256 & 258 & 66306 & 28.39 & 9.56 & 0.59 & 16.59 & 3.17 & 6.76\\
PUC-512 & 514 & 263682 & 34.50 & 73.28 & 0.00 & 24.43 & 30.72 & 0.00\\
PUC-1024 & 1026 & 1051650 & 43.67 & 1378.66 & 0.00 & 28.99 & 464.61 & 0.00\\
\hline
\end{tabular}}
\end{table}

From these results, we observe that the dual ascent requires a very limited CPU time on the majority of instances. The random criterion produces better dual bounds than the greedy-cut selection, in less CPU time, and was therefore selected for the remainder of our experiments. The solutions of the dual ascent help fixing $78.30\%$ of the edges in average, therefore diminishing the number of variables considered in the B\&C. This is a significant help, given that $G$ is a complete graph and the current instances can lead to time-consuming LPs with up to one million variables. Even so, in contrast with the Steiner-tree problem \citep{de2001dual}, dual ascent bounds tend to deteriorate on large instances, thus limiting possible problem simplifications in these situations. This deterioration seems to be a feature of the model, since the root linear relaxations reported later in this section follow the same trend. This difference of behavior can be connected to the fact that the MSFBC involves a  \emph{forest} of trees rather than a \emph{single} one.

\subsubsection{Branch-and-cut}
\label{section:BC}

Table \ref{tab:bc_sum} summarizes the results of our B\&C algorithm. The time limit is set to 3600 seconds. We report the following information, averaged over each group: 
\begin{itemize}[leftmargin=*]
\item \textbf{GAP$_\textsc{Root}$(\%)}: the percentage gap between the BKS and the root node LP relaxation;
\item \textbf{GAP$_\textsc{Final}$(\%)}: the percentage gap at the end of the B\&C;
\item \textbf{OPT}: the number of instances in the group solved to optimality;
\item \textbf{T(s)}: the CPU time of the algorithm, in seconds;
\item \textbf{T$_\textsc{Flow}$(s)}: the CPU time spent in the max-flow/min-cut algorithm, for cut separation;
\item \textbf{T$_\textsc{Root}$(s)}: the CPU time solving the root node.
\item \textbf{N$_\textsc{Node}$}: the number of nodes explored in the branch-and-bound tree.
\item \textbf{N$_\textsc{Tree}$}: the number of trees in the optimal solution ;
\item \textbf{S$_\textsc{Tree}$}: the average size of the trees in the optimal solution.
\end{itemize}

\begin{table}[htbp]
\centering
\renewcommand{\arraystretch}{1.2}
\setlength{\tabcolsep}{6.5pt}
\caption{Results of the branch-and-cut algorithm}
\label{tab:bc_sum}
\scalebox{0.8}
{
\begin{tabular}{|c|ccc|ccc|ccc|}
\hline
Group & GAP$_\textsc{Root}$(\%) & GAP$_\textsc{Final}$(\%) & OPT & T(s) & T$_\textsc{Flow}$(s) & T$_\textsc{Root}$(s) & N$_\textsc{Node}$ & N$_\textsc{Tree}$ & S$_\textsc{Tree}$ \\ \hline 
PUC-8 & 0.00 & 0.00 & 5/5 & $<$0.01 & $<$0.01 & $<$0.01 & 1.00 & 2.20 & 4.67 \\
PUC-12 & 0.00 & 0.00 & 5/5 & $<$0.01 & $<$0.01 & $<$0.01 & 1.00 & 3.60 & 4.06 \\
PUC-16 & 0.00 & 0.00 & 5/5 & $<$0.01 & $<$0.01 & $<$0.01 & 1.00 & 4.80 & 4.17  \\ 
PUC-20 & 1.20 & 0.00 & 5/5 & $<$0.01 & $<$0.01 & $<$0.01 & 1.00 & 6.00 & 3.89 \\
PUC-24 & 1.28 & 0.00 & 5/5 & $<$0.01 & $<$0.01 & $<$0.01 & 1.00 & 7.40 & 3.63 \\
PUC-28 & 1.94 & 0.00 & 5/5 & $<$0.01 & $<$0.01 & $<$0.01 & 1.00 & 7.60 & 4.06 \\ 
PUC-32 & 2.58 & 0.00 & 5/5 & 0.01 & $<$0.01 & 0.01 & 1.00 & 10.60 & 3.28 \\ 
PUC-36 & 3.87 & 0.00 & 5/5 & $<$0.01 & $<$0.01 & $<$0.01 & 1.00 & 10.00 & 3.88 \\ 
PUC-40 & 3.21 & 0.00 & 5/5 & 2.34 & 0.19 & 0.05 & 11.80 & 12.60 & 3.42 \\ 
PUC-44 & 3.84 & 0.00 & 5/5 & 4.78 & 0.75 & 0.05 & 23.40 & 13.80 & 3.44 \\
PUC-48 & 2.95 & 0.00 & 5/5 & $<$0.01 & $<$0.01 & $<$0.01 & 1.00 & 15.60 & 3.36 \\ 
PUC-52 & 4.53 & 0.00 & 5/5 & $<$0.01 & $<$0.01 & $<$0.01 & 1.00 & 18.80 & 2.89 \\
PUC-56 & 4.22 & 0.00 & 5/5 & 0.02 & $<$0.01 & 0.02 & 1.00 & 17.20 & 3.55 \\
PUC-60 & 2.58 & 0.00 & 5/5 & 0.32 & 0.05 & 0.08 & 3.00 & 19.60 & 3.19 \\ 
PUC-64 & 2.27 & 0.00 & 5/5 & 3.21 & 0.21 & 0.18 & 10.20 & 12.80 & 13.91 \\ 
PUC-80 & 1.74 & 0.00 & 5/5 & 93.05 & 8.51 & 0.21 & 319.80 & 24.60 & 7.55 \\ 
PUC-96 & 0.92 & 0.00 & 5/5 & 461.51 & 39.40 & 0.84 & 613.00 & 25.00 & 8.67 \\
PUC-128 & 1.45 & 0.00 & 5/5 & 1986.37 & 202.07 & 1.07 & 1525.40 & 44.40 & 2.96 \\
PUC-256 & 9.15 & 1.83 & 0/5 & 3600.00 & 342.87 & 27.24 & 167.00 & -- & -- \\
PUC-512 & 16.66 & 16.43 & 0/5 & 3600.00 & 262.90 & 738.28 & 6.00 & -- & -- \\ 
PUC-1024 & 21.06 & 21.06 & 0/5 & 3600.00 & -- & $>$3600$^\dagger$ & 1.00 & -- & -- \\
\hline 
\end{tabular}
}

\vspace*{0.2cm}
\begin{flushleft}
\scriptsize
\ \ $\dagger$ -- The solution of the root node was not completed
\end{flushleft}
\end{table}

From these experiments, we observe that the B\&C can currently solve all instances with up to 128 residue vertices, for a total of 90 instances out of the 105 available. For the remaining 15 instances, an average optimality gap of $13.10\%$ is obtained after 3600 seconds. 

The distribution of the CPU currently indicates that the cut-separation procedure is efficient, as the time spent in the max-flow algorithm occupies only a small fraction of the overall solution time, and remains inferior to the time spent solving the LPs. The computation of the lower bound can be time consuming for large problems, due to the large number of variables and the medium to high density of the constraint matrix as a consequence of the cut separation. In particular, for the largest problem instances, the root node solution was not completed in the allowed time.

Finally, the results of the exact method contributed to validate the solution quality of the metaheuristic presented in the previous section, which retrieved 80 out of the 90 known optimal solutions. Based on these results, the metaheuristic approach is more adequate for large-scale problems with several hundreds or thousands of residues. The following section investigates the performance of this method on data sets issued from interferometric synthetic-aperture radar (InSAR) and magnetic resonance imaging (MRI) applications.

\subsection{Application to 2D phase unwrapping}

We now investigate the performance of our approach on the 2D phase unwrapping application. Four metrics can be used to compare the quality of each solution with those of other path-following methods: 
\begin{itemize}
\item[--] \textbf{N}: the number of absolute phase gradients that differ from their wrapped counterparts;
\item[--] \textbf{L}: the total length of the branch-cuts; 
\item[--] \textbf{T}: the number of trees produced by the branch-cuts;
\item[--] \textbf{I}: the number of isolated regions enclosed by branch-cuts and/or residues. \\
\end{itemize}

We conducted our experiments on a classical set of benchmark instances, introduced by \cite{ghiglia1998two}: \emph{Long's Peak}, \emph{Isola's Peak}, and \emph{Head Magnetic Resonance Image}. Each instance poses different challenges. 
 
To evaluate our approach, we consider the primal solution produced by our HILS metaheuristic with a time limit of 3600 seconds. The first two instances have additional information that can mask regions with poor-quality pixels. Our methods used this information to limit the relevant area for the unwrapping process, and not to redefine the positions of the border points. We compare our results with those of two seminal path-following methods: Goldstein's algorithm \citep{goldstein1988satellite} and Buckland's minimum-cost matching algorithm \citep{buckland1995unwrapping}.

\subsubsection{Long's Peak}

\begin{figure}[htbp]
    \centering
    \subfigure[]{\label{fig:longs_w}\fbox{\includegraphics[scale=0.3]{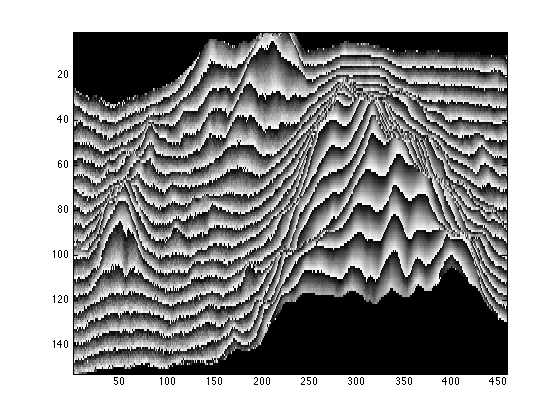}}}
    \hspace{1em}
     \subfigure[]{\label{fig:longs_res}\fbox{\includegraphics[scale=0.3]{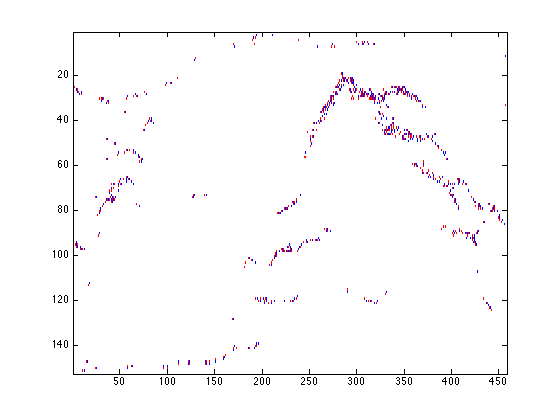}}}
    \hspace{1em}
     \subfigure[]{\label{fig:longs_gold}\fbox{\includegraphics[scale=0.3]{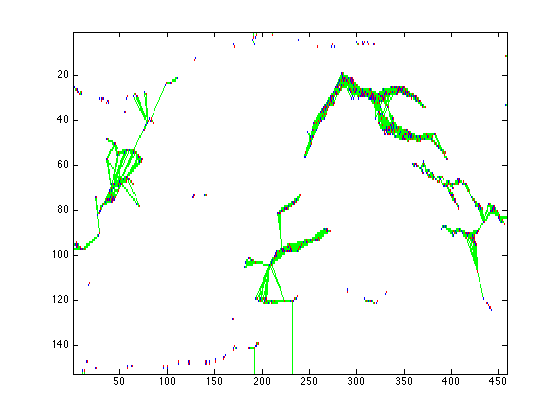}}}
    \hspace{1em}
     \subfigure[]{\label{fig:longs_gold_unwr}\fbox{\includegraphics[scale=0.3]{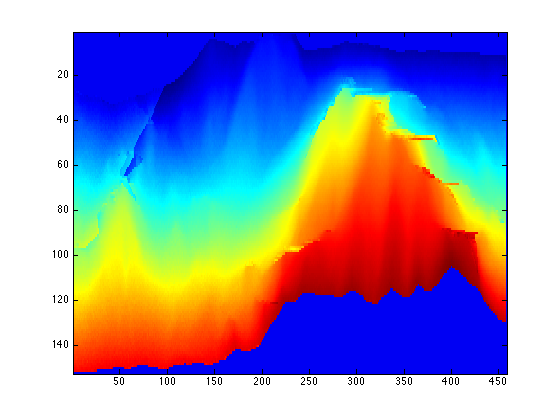}}}
    \hspace{1em}
     \subfigure[]{\label{fig:longs_hung}\fbox{\includegraphics[scale=0.3]{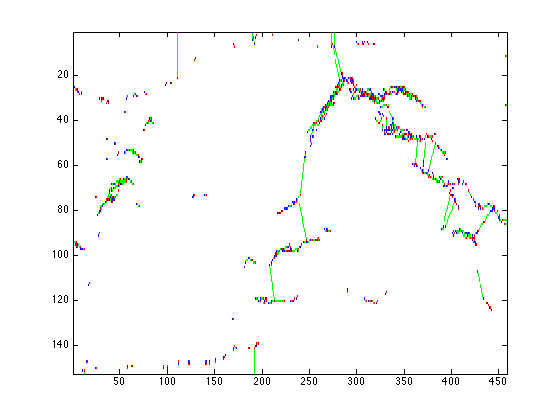}}}
    \hspace{1em}
     \subfigure[]{\label{fig:longs_hung_unwr}\fbox{\includegraphics[scale=0.3]{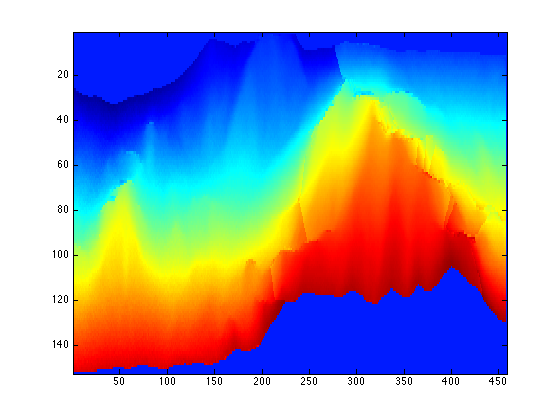}}}
    \hspace{1em}
     \subfigure[]{\label{fig:longs_msfbc}\fbox{\includegraphics[scale=0.3]{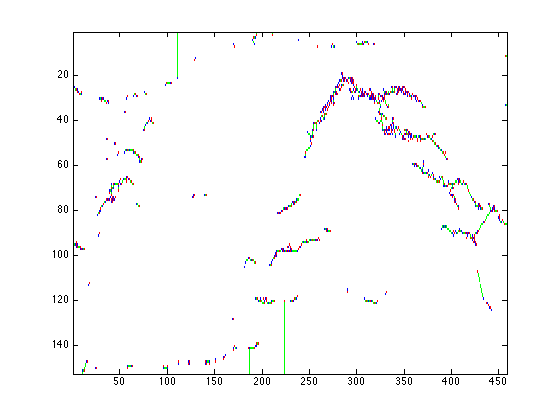}}}
    \hspace{1em}
    \subfigure[]{\label{fig:longs_msfbc_unwr}\fbox{\includegraphics[scale=0.3]{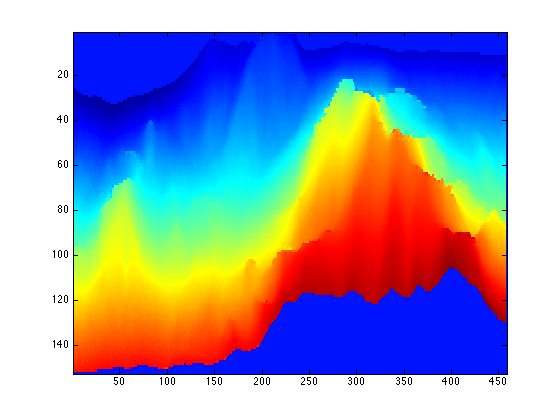}}}
    \caption{Long's Peak instance with 846 residues distributed over a 152$\times$458-pixel image. Results of Goldstein's algorithm, MCM algorithm, and MSFBCP approach. (a) Wrapped phase image. (b) Topology of residues. (c) Goldstein's branch-cut configuration. (d) Goldstein's unwrapped solution. (e) MCM branch-cut configuration. (f) MCM unwrapped solution. (g) MSFBCP branch-cut configuration. (h) MSFBCP unwrapped solution.}
    \label{fig:longs}
\end{figure}

Figure \ref{fig:longs}(a) shows the wrapped phase image of a steep mountainous region in Colorado (USA) called Long's Peak, obtained from a high-fidelity InSAR simulator \citep{ghiglia1998two}. The topology of the residues clearly suggests natural phase discontinuities caused by the terrain's elevation as well as the presence of noise. There are 846 residues (422 positives and 426 negatives) distributed over a 152$\times$458-pixel image. The greatest challenge is to efficiently cluster the sparse group of residues. Many regions have a high density of residues, which could lead to isolated regions in the solution. 

Figures \ref{fig:longs}(c) and \ref{fig:longs}(d) show that although Goldstein's algorithm was able to cluster groups of residues, the number of unnecessary long connections and isolated regions had a negative impact on the unwrapping result. The minimum cost matching (MCM) algorithm (Figures~\ref{fig:longs}(e) and \ref{fig:longs}(f)) failed to preserve the structural boundaries suggested by the topology of the residues, introducing visible discontinuities. On the other hand, the MSFBCP approach was able to efficiently cluster groups of residues and respect their topology, and it greatly reduced the number of discontinuities introduced by the branch-cuts. Table~\ref{tab:longs_t} gives the solution metrics for the three approaches.

%%%%%%%%%%%%%%%%%%%%%%%%%%%%%%%%%%%%%%%%%%%%%%%%%%%%%%%%
\begin{table}[htbp]
\centering
\caption{Results for Long's Peak data set}
\label{tab:longs_t}
\renewcommand{\arraystretch}{1.2}
\begin{tabular}{|c|c|c|c|c|}
\hline
Method & N & L & T & I \\ \hline 
Goldstein & 1437 & 10647.96 & 49 & 110 \\
MCM & 1075 & 1545.38 & 429 & 47 \\
MSFBCP & 975 & 1264.31 & 68 & 25 \\ \hline
\end{tabular}
\end{table}
%%%%%%%%%%%%%%%%%%%%%%%%%%%%%%%%%%%%%%%%%%%%%%%%%%

The MSFBCP has 32.15\% fewer discontinuity points than Goldstein's algorithm, and 9.30\% fewer than the MCM algorithm. In comparison with the MCM, the total length of the branch-cuts is improved by 18\% and the number of isolated regions by 47\%. 

\subsubsection{Isola's Peak}

\begin{figure}[htbp]
    \centering
    \subfigure[]{\label{fig:isola_w}\fbox{\includegraphics[scale=0.3]{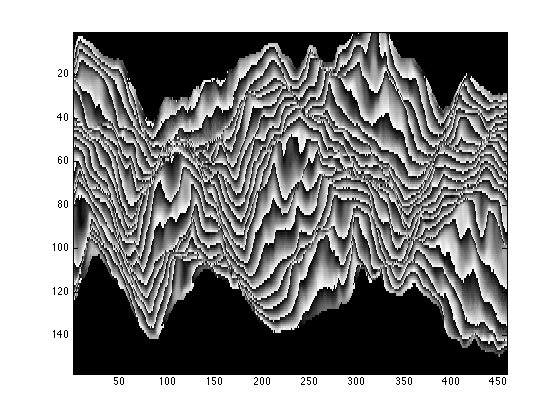}}}
    \hspace{1em}
     \subfigure[]{\label{fig:isola_res}\fbox{\includegraphics[scale=0.3]{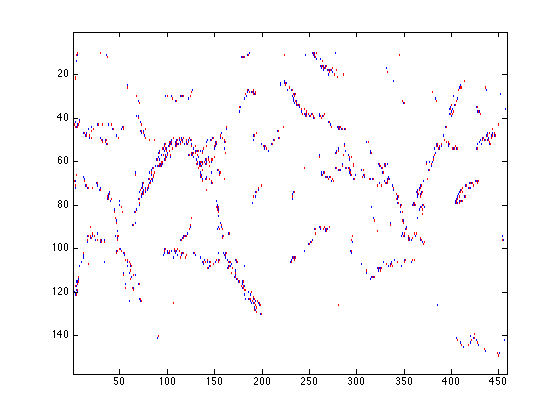}}}
    \hspace{1em}
     \subfigure[]{\label{fig:isola_gold}\fbox{\includegraphics[scale=0.3]{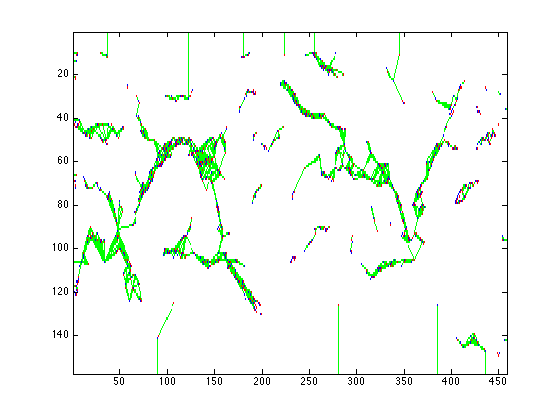}}}
    \hspace{1em}
     \subfigure[]{\label{fig:isola_gold_unwr}\fbox{\includegraphics[scale=0.3]{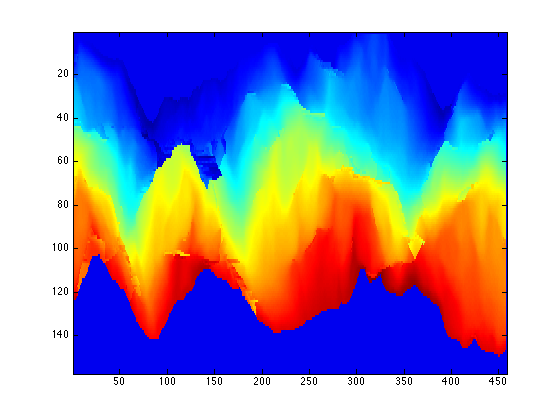}}}
    \hspace{1em}
     \subfigure[]{\label{fig:isola_hung}\fbox{\includegraphics[scale=0.3]{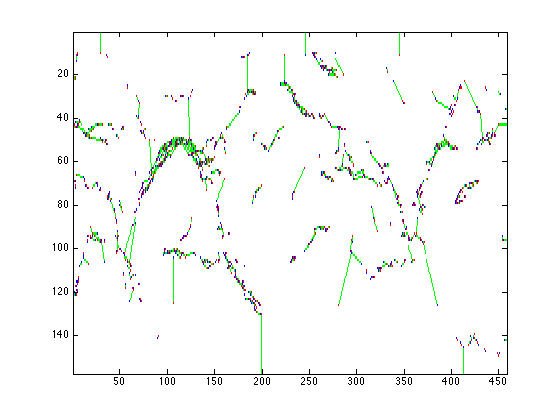}}}
    \hspace{1em}
     \subfigure[]{\label{fig:isola_hung_unwr}\fbox{\includegraphics[scale=0.3]{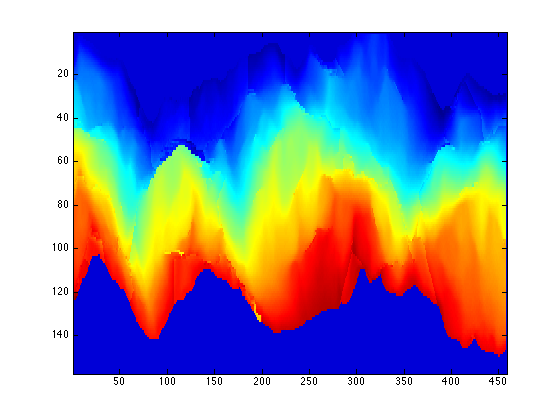}}}
    \hspace{1em}
     \subfigure[]{\label{fig:isola_msfbc}\fbox{\includegraphics[scale=0.3]{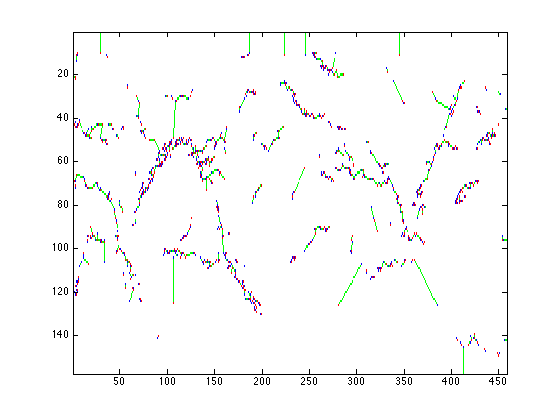}}}
    \hspace{1em}
    \subfigure[]{\label{fig:isola_msfbc_unwr}\fbox{\includegraphics[scale=0.3]{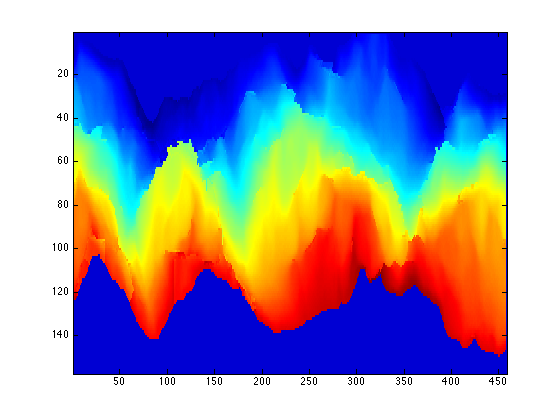}}}
    \caption{Isola's Peak instance with 1234 residues distributed over a 157$\times$458-pixel image. Results of Goldstein's algorithm, MCM algorithm, and MSFBCP approach. (a) Wrapped phase image. (b) Topology of residues. (c) Goldstein's branch-cut configuration. (d) Goldstein's unwrapped solution. (e) MCM branch-cut configuration. (f) MCM unwrapped solution. (g) MSFBCP branch-cut configuration. (h) MSFBCP unwrapped solution.}
    \label{fig:isola}
\end{figure}

Figure \ref{fig:isola}(a) shows the wrapped phase image of a steep-relief mountainous region in Colorado (USA) called Isola's Peak, obtained from a high-fidelity InSAR simulator. The challenge here is to correctly unwrap the phase data around the numerous regions with natural phase discontinuities, without propagating errors on the unwrapped surface. There are 1234 residues (616 positives and 618 negatives) distributed over a 157$\times$458-pixel image. Again, the goals are to efficiently cluster the sparse groups of residues and to handle the high-density regions.

As Figure~\ref{fig:isola} shows, the branch-cuts produced by the MSFBCP approach are less visible in the image (Figure \ref{fig:isola}(g)). The groups of residues are better clustered, and most structural boundaries are respected. Table~\ref{tab:isola_t} gives the solution metrics for the three approaches.

%%%%%%%%%%%%%%%%%%%%%%%%%%%%%%%%%%%%%%%%%%%%%%%%%%%%%%%%
\begin{table}[htbp]
\centering
\caption{Results for Isola's Peak data set}
\label{tab:isola_t}
\renewcommand{\arraystretch}{1.2}
\begin{tabular}{|c|c|c|c|c|}
\hline
Method & N & L & T & I \\ \hline 
Goldstein & 2127 & 11578.36 & 39 & 226\\
MCM & 1825 & 2545.06 & 625 & 85 \\
MSFBCP & 1609 & 1850.23 & 57 & 29 \\ \hline
\end{tabular}
\end{table}
%%%%%%%%%%%%%%%%%%%%%%%%%%%%%%%%%%%%%%%%%%%%%%%%%%

The MSFBCP solution has 24.35\% fewer discontinuity points than Goldstein's algorithm, and 11.83\% fewer than the MCM algorithm. In comparison with the MCM, the total length of the branch-cuts is improved by 27.3\% and the number of isolated regions by 65.8\%. 

\subsubsection{Head MRI.}

\begin{figure}[htbp]
    \centering
    \subfigure[]{\label{fig:head_w}\fbox{\includegraphics[scale=0.3]{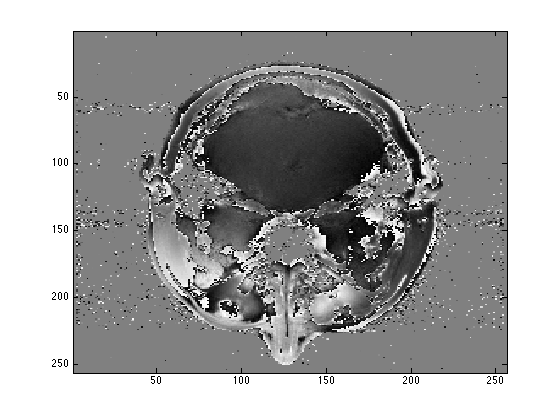}}}
    \hspace{1em}
     \subfigure[]{\label{fig: head_res}\fbox{\includegraphics[scale=0.3]{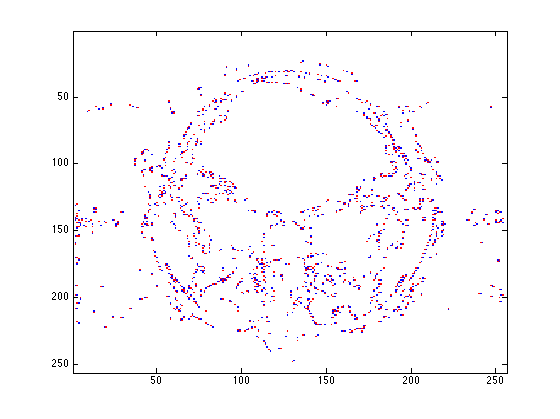}}}
    \hspace{1em}
     \subfigure[]{\label{fig:head_gold}\fbox{\includegraphics[scale=0.3]{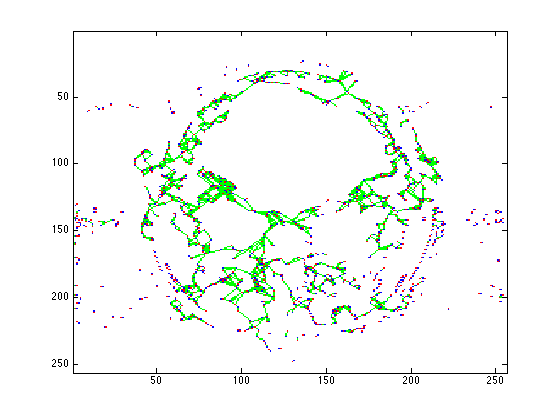}}}
    \hspace{1em}
     \subfigure[]{\label{fig:head_gold_unwr}\fbox{\includegraphics[scale=0.3]{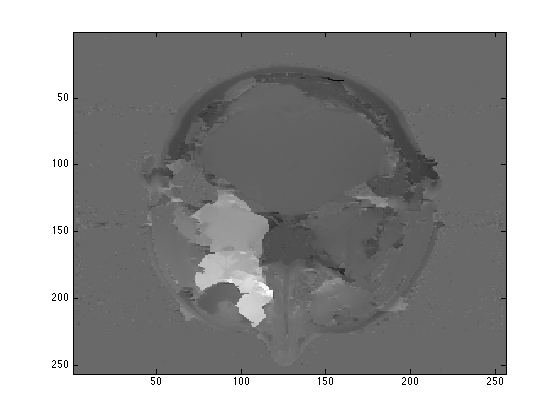}}}
    \hspace{1em}
     \subfigure[]{\label{fig: head_hung}\fbox{\includegraphics[scale=0.3]{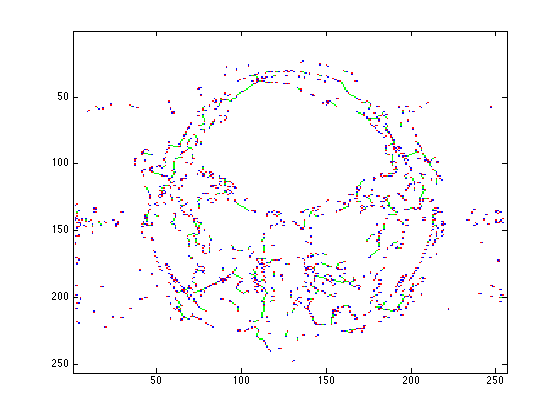}}}
    \hspace{1em}
     \subfigure[]{\label{fig: head_hung_unwr}\fbox{\includegraphics[scale=0.3]{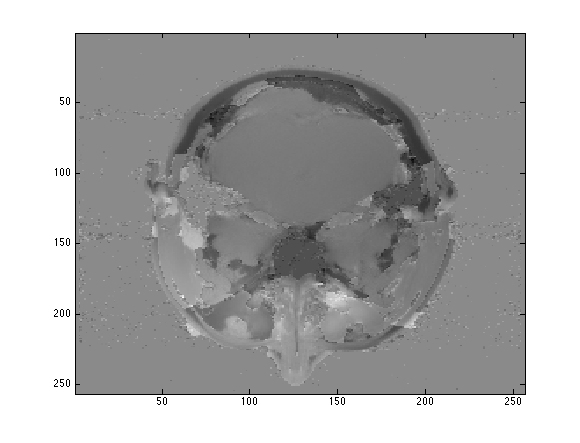}}}
    \hspace{1em}
     \subfigure[]{\label{fig: head_msfbc}\fbox{\includegraphics[scale=0.3]{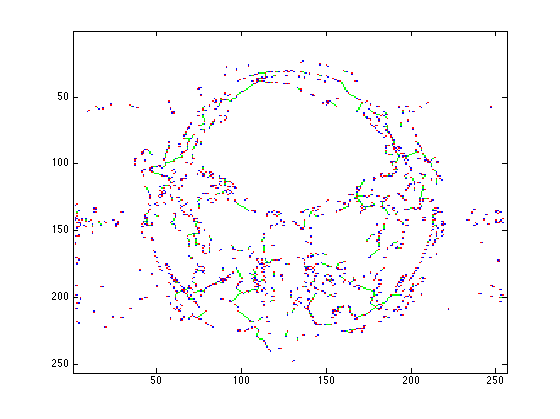}}}
    \hspace{1em}
    \subfigure[]{\label{fig: head_msfbc_unwr}\fbox{\includegraphics[scale=0.3]{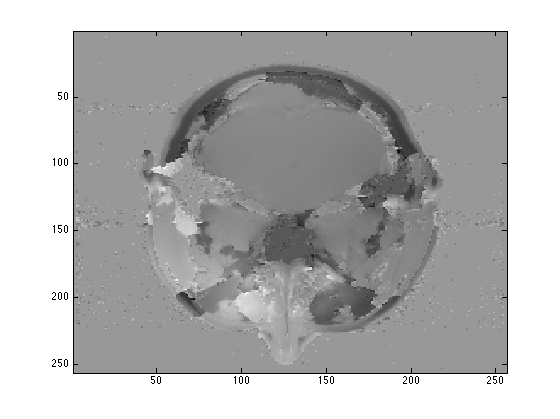}}}
    \caption{Head MRI instance with 1926 residues distributed over a 256$\times$256-pixel image. Results of Goldstein's algorithm, MCM algorithm, and MSFBCP approach. (a) Wrapped phase image. (b) Topology of residues. (c) Goldstein's branch-cut configuration. (d) Goldstein's unwrapped solution. (e) MCM branch-cut configuration. (f) MCM unwrapped solution. (g) MSFBCP branch-cut configuration. (h) MSFBCP unwrapped solution.}
    \label{fig:head}
\end{figure}

Figure~\ref{fig:head}(a) shows a wrapped phase image of a head MRI with 1926 residues (963 positives and 963 negatives) defined on a 256$\times$256-pixel grid. The signal obtained from conventional MRI procedures captures the water and fat intensity values from tissues, shifted by a phase value in each pixel. The continuous phase signal is needed to segment the water and fat areas of the image. This is a difficult instance, since various regions appear at first glance to be almost isolated.

As illustrated in Figure~\ref{fig:head}(c), Goldstein's algorithm failed to cluster the groups of residues, creating many long connections and isolated regions. The outcome of such erroneous branch-cut placements is apparent (Figure~\ref{fig:head}(d)). The MCM and MSFBCP approaches produced much better solutions. Indeed, most of the residues appear as close dipoles, and thus we expect the MCM solution to be a very good approximation of the  L$^{0}$--norm problem. Table~\ref{tab:head_t} gives the solution metrics for the three approaches.

%%%%%%%%%%%%%%%%%
%%%%%%%%%%%%%%%%%%%%%%%%%%%%%%%%%%%%%%%%%%%%%%%%%%%%%%%%%%%%%%%%%%%%%%

%%%%%%%%%%%%%%%%%
%%%%%%%%%%%%%%%%%%%%%%%%%%%%%%%%%%%%%%%%%%%%%%%%%%%%%%%%%%%%%%%%%%%%%%

%%%%%%%%%%%%%%%%%%%%%%%%%%%%%%%%%%%%%%%%%%%%%%%%%%%%%%%%%%%%%%%%%%%%%%%%
%%%%%%%%%%%%%%%%%%%%%%%%%%%%%%%%%%%%%%%%%%%%%%%%%%%%%%%%%%%%%%%%%%%%%%

%%%%%%%%%%%%%%%%%%%%%%%%%%%%%%%%%%%%%%%%%%%%%%%%%%%%%%%%
\begin{table}[htbp]
\centering
\caption{Results for Head MRI data set}
\label{tab:head_t}
\renewcommand{\arraystretch}{1.2}
\begin{tabular}{|c|c|c|c|c|}
\hline
Method & N & L & T & I \\ \hline 
Goldstein & 2570 & 11696.44 & 153 & 257\\
MCM & 1789 & 1588.72 & 963 & 16 \\
MSFBCP & 1810 & 1722.56 & 57 & 19 \\ \hline
\end{tabular}
\end{table}
%%%%%%%%%%%%%%%%%%%%%%%%%%%%%%%%%%%%%%%%%%%%%%%%%%

The MSFBCP solution has 29.87\% fewer discontinuity points than Goldstein's algorithm, and 1.16\% more than the MCM algorithm. 
In comparison with the MCM, the total length of the branch-cuts is 8.42\% higher, and there are three more isolated regions.
The MCM solution is likely to be a very good approximation of the optimal L$^{0}$--norm solution for this data set, since it includes many close residue pairs.
Note that the MSFBCP formulation includes the MCM solution in its feasible solution space. Hence, an optimal solution is guaranteed to have shorter branch-cuts than previous approaches. However, our metaheuristics may lead to an optimality gap for large instances, as observed for this data set. In future research, we could investigate using the MCM algorithm to construct an initial solution, hence guaranteeing an improvement in every case.

\section{Conclusions}

We have proposed an MSFBCP formulation of the L$^{0}$-norm 2D phase unwrapping problem, as well as efficient combinatorial optimization algorithms to solve it. Our B\&C algorithm successfully solves all instances with up to 128 residues in less than one hour of CPU time. To solve larger problems, we introduced a hybrid metaheuristic based on iterated local search with a set partitioning formulation. The exact and heuristic solutions are identical on 80 of the 90 known optima.

To evaluate the MSFBCP solutions in the context of a 2D phase unwrapping application, we considered three wrapped images from InSAR and MRI applications. Our approach consistently produced high-quality images with few visible discontinuities. When solved to optimality, the MSFBCP model is guaranteed to produce branch-cuts that are shorter than those of the previous algorithms, and we observed significant quality improvements on two of the three instances.
For the third instance, with a residue configuration forming natural pairs, the MCM algorithm performed slightly better because our metaheuristic did not find the optimal solution. Future work for the MSFBCP will develop ways to better exploit this new formulation. Overall, phase unwrapping leads to challenging and interesting mathematical problems with applications in several areas of computer graphics. Work at the frontier between signal processing and operations research can lead to exciting new developments in the solution of such problems.

\ACKNOWLEDGMENT{
This research is partially supported by CNPq, CAPES and FAPERJ in Brazil. This support is gratefully acknowledged. The authors wish to thank the two anonymous referees for their detailed reports, which significantly contributed to improve this paper.
}

%\bibliographystyle{informs2014}
%\bibliography{libraryJOC}

\begin{thebibliography}{21}
\providecommand{\natexlab}[1]{#1}
\providecommand{\url}[1]{\texttt{#1}}
\providecommand{\urlprefix}{URL }

\bibitem[{Buckland et~al.(1995)Buckland, Huntley, \protect\BIBand{}
  Turner}]{buckland1995unwrapping}
Buckland J, Huntley J, Turner S (1995) Unwrapping noisy phase maps by use of a
  minimum-cost-matching algorithm. \emph{Applied Optics} 34(23):5100--5108.

\bibitem[{Chen \protect\BIBand{} Zebker(2000)}]{chen2000network}
Chen CW, Zebker HA (2000) Network approaches to two-dimensional phase
  unwrapping: intractability and two new algorithms. \emph{JOSA A}
  17(3):401--414.

\bibitem[{Chen \protect\BIBand{} Zebker(2001)}]{chen2001two}
Chen CW, Zebker HA (2001) Two-dimensional phase unwrapping with use of
  statistical models for cost functions in nonlinear optimization. \emph{JOSA
  A} 18(2):338--351.

\bibitem[{Claus \protect\BIBand{} Maculan(1983)}]{claus1983nouvelle}
Claus A, Maculan N (1983) Une nouvelle formulation du probleme de {S}teiner sur
  un graphe. Technical report, Montr{\'e}al: Universit{\'e} de Montr{\'e}al,
  Centre de recherche sur les transports.

\bibitem[{Curlander \protect\BIBand{} McDonough(1991)}]{curlander1991synthetic}
Curlander JC, McDonough RN (1991) \emph{Synthetic Aperture Radar} (John Wiley
  \& Sons, New York).

\bibitem[{de~Arag{\~a}o et~al.(2001)de~Arag{\~a}o, Uchoa, \protect\BIBand{}
  Werneck}]{de2001dual}
de~Arag{\~a}o MP, Uchoa E, Werneck RF (2001) Dual heuristics on the exact
  solution of large {S}teiner problems. \emph{Electronic Notes in Discrete
  Mathematics} 7:150--153.

\bibitem[{Ghiglia et~al.(1987)Ghiglia, Mastin, \protect\BIBand{}
  Romero}]{ghiglia1987cellular}
Ghiglia DC, Mastin GA, Romero LA (1987) Cellular-automata method for phase
  unwrapping. \emph{JOSA A} 4(1):267--280.

\bibitem[{Ghiglia \protect\BIBand{} Pritt(1998)}]{ghiglia1998two}
Ghiglia DC, Pritt MD (1998) \emph{Two-dimensional Phase Unwrapping: Theory,
  Algorithms, and Software} (Wiley, New York).

\bibitem[{Ghiglia \protect\BIBand{} Romero(1996)}]{ghiglia1996minimum}
Ghiglia DC, Romero LA (1996) Minimum ${L}^p$-norm two-dimensional phase
  unwrapping. \emph{JOSA A} 13(10):1999--2013.

\bibitem[{Glover \protect\BIBand{} Schneider(1991)}]{glover1991three}
Glover G, Schneider E (1991) Three-point dixon technique for true water/fat
  decomposition with ${B}_0$ inhomogeneity correction. \emph{Magnetic Resonance
  in Medicine} 18(2):371--383.

\bibitem[{Goldberg \protect\BIBand{} Tarjan(1988)}]{goldberg1988new}
Goldberg AV, Tarjan RE (1988) A new approach to the maximum-flow problem.
  \emph{Journal of the ACM} 35(4):921--940.

\bibitem[{Goldstein et~al.(1988)Goldstein, Zebker, \protect\BIBand{}
  Werner}]{goldstein1988satellite}
Goldstein RM, Zebker HA, Werner CL (1988) Satellite radar interferometry:
  Two-dimensional phase unwrapping. \emph{Radio Science} 23(4):713--720.

\bibitem[{Guizar-Sicairos et~al.(2011)Guizar-Sicairos, Diaz, Holler, Lucas,
  Menzel, Wepf, \protect\BIBand{} Bunk}]{guizar2011phase}
Guizar-Sicairos M, Diaz A, Holler M, Lucas MS, Menzel A, Wepf RA, Bunk O (2011)
  Phase tomography from x-ray coherent diffractive imaging projections.
  \emph{Optics Express} 19(22):21345--21357.

\bibitem[{Huntley \protect\BIBand{}
  Buckland(1995)}]{huntley1995characterization}
Huntley J, Buckland J (1995) Characterization of sources of 2$\pi$ phase
  discontinuity in speckle interferograms. \emph{JOSA A} 12(9):1990--1996.

\bibitem[{Itoh(1982)}]{itoh1982analysis}
Itoh K (1982) Analysis of the phase unwrapping algorithm. \emph{Applied Optics}
  21(14):2470--2470.

\bibitem[{Miller(2001)}]{miller2001comparison}
Miller W (2001) Comparison of genomic {DNA} sequences: Solved and unsolved
  problems. \emph{Bioinformatics} 17(5):391--397.

\bibitem[{Pandit et~al.(1994)Pandit, Jordache, \protect\BIBand{}
  Joshi}]{Pandit1994}
Pandit S, Jordache N, Joshi G (1994) {Data-dependent systems methodology for
  noise-insensitive phase unwrapping in laser interferometric surface
  characterization}. \emph{Journal of the Optical Society of America A}
  11(10):2584--2592.

\bibitem[{Sawaf \protect\BIBand{} Tatam(2006)}]{sawaf2006finding}
Sawaf F, Tatam RP (2006) Finding minimum spanning trees more efficiently for
  tile-based phase unwrapping. \emph{Measurement Science and Technology}
  17(6):1428--1435.

\bibitem[{Swofford et~al.(1990)Swofford, Olsen, Waddell, \protect\BIBand{}
  Hillis}]{swofford1990phylogeny}
Swofford DL, Olsen GJ, Waddell P, Hillis D (1990) Phylogeny reconstruction.
  \emph{Molecular Systematics} 3:411--501.

\bibitem[{Wong(1984)}]{wong1984dual}
Wong RT (1984) A dual ascent approach for {S}teiner tree problems on a directed
  graph. \emph{Mathematical Programming} 28(3):271--287.

\bibitem[{Zebker \protect\BIBand{} Lu(1998)}]{zebker1998phase}
Zebker HA, Lu Y (1998) Phase unwrapping algorithms for radar interferometry:
  Residue-cut, least-squares, and synthesis algorithms. \emph{JOSA A}
  15(3):586--598.

\end{thebibliography}

\ECSwitch

\SingleSpacedXI

\ECHead{\centering 2D-Phase Unwrapping via Balanced Spanning Forests \linebreak \emph{Electronic Companion}}

\vspace*{1cm}

\begin{center}
\textbf{Ian Herszterg$^a$, Marcus Poggi$^b$, Thibaut Vidal$^b$} \\ \vspace*{0.4cm}

$^a$ School of Industrial and Systems Engineering, Georgia Institute of Technology, USA \\
iherszterg@gatech.edu \\
$^b$ Departamento de Inform\'{a}tica, Pontif\'{i}cia Universidade Cat\'{o}lica do Rio de Janeiro, Brazil \\
\{poggi,vidalt\}@inf.puc-rio.br \\
\end{center}

\vspace*{0.6cm}

\section*{The MSFBCP is $\mathcal{NP}$-hard}

\emph{Proof:} Let $G = (V,E)$ be a graph with positive edge costs and a set of terminal vertices $T \subseteq V$. The Steiner-tree problem seeks to find a 
minimum-cost connected subgraph $G' = (V',E')$ with $T \subseteq V'$. We can reduce an instance of the Steiner problem in graphs to an MSFBCP instance by assigning a positive weight of $w = 1$ to all the terminal vertices, except for one, which is replaced by $\abs{T}-1$ vertices with a negative weight of $w = -1$, connected to each other with cost zero. In addition, a pair of connected vertices with opposite signed weights is located for each of the $(V-T)$ remaining vertices (Steiner points). The cost of connecting such a pair of vertices is also set to zero. 

%%%%%%%%%%%%%%%%%%%%%%%%%%%%%%%%%%%%%%%%%%%%%%%%%%%%%%%%%%%%%%%%%%%%%%%%
\begin{figure}[htbp]
    \centering
    \subfigure[]{\label{fig:steiner_inst}\fbox{\includegraphics[scale=0.42]{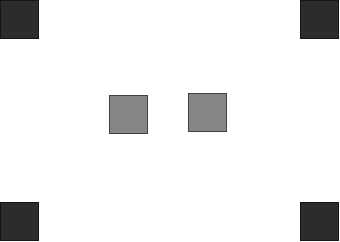}}}
    \hspace{1em}
    \subfigure[]{\label{fig:msfbc}\fbox{\includegraphics[scale=0.42]{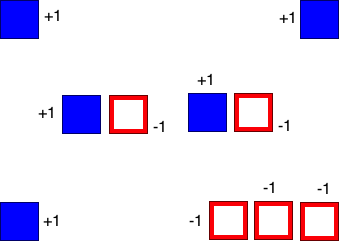}}}
    \caption{ \emph{Reduction of the Steiner-tree problem to the MSFBCP: (a) Steiner-tree instance, with four terminal vertices (black) and two Steiner-points (grey). (b) Equivalent instance for the MSFBCP. Positive residues are represented with solid blue squares, and negative residues are represented with empty red squares.}\label{fig:steiner_reduc}}
\end{figure}
%%%%%%%%%%%%%%%%%%%%%%%%%%%%%%%%%%%%%%%%%%%%%%%%%%%%%%%%%%%%%%%%%%%%%%

\begin{figure}[htbp]
    \centering
    \subfigure[]{\label{fig:steiner_sol}\fbox{\includegraphics[scale=0.4]{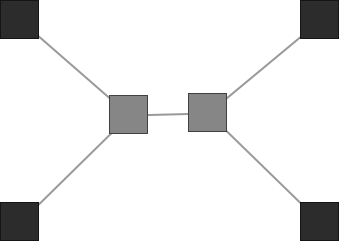}}}
    \hspace{5em}
    \subfigure[]{\label{fig:msfbc_optimal}\fbox{\includegraphics[scale=0.4]{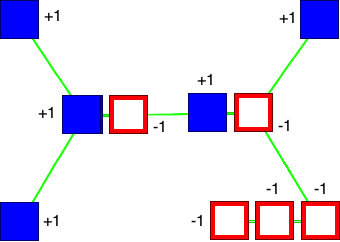}}}
        \hspace{5em}
    \subfigure[]{\label{fig:msfbc_feasible}\fbox{\includegraphics[scale=0.4]{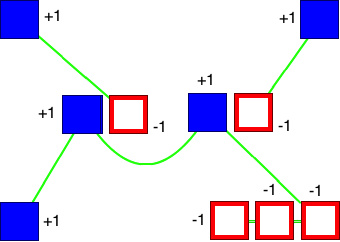}}}
     \hspace{5em}
    \subfigure[]{\label{fig:msfbc_infeasible}\fbox{\includegraphics[scale=0.4]{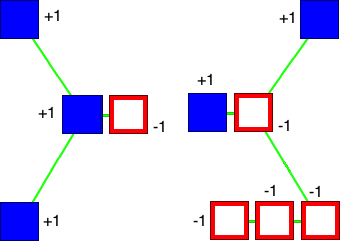}}}
    \caption{Optimal solution for the Steiner-tree problem and two feasible solutions for the MSFBCP: (a) Optimal solution for the Steiner-tree problem. (b) An optimal MSFBCP solution that is also feasible and optimal for the Steiner-tree problem. (c) An MSFBCP solution with the same cost that is infeasible for the Steiner-tree problem. (d) An infeasible MSFBCP solution.}
    \label{fig:reduc_sols}
\end{figure}

Figure \ref{fig:steiner_reduc} shows the reduction of a Steiner-tree instance with four terminal nodes and two Steiner nodes, while Figure \ref{fig:steiner_sol} shows the optimal solution for the Steiner-tree problem. In the MSFBCP solution of the reduced instance, one of the following cases occurs: 
\begin{enumerate}[nosep]
\item  A single balanced tree spans all the terminal vertices (Figure \ref{fig:msfbc_optimal});
\item  At least two disjoint balanced trees contain the terminal vertices (Figure \ref{fig:msfbc_feasible}).
\end{enumerate}

In the first case, the corresponding minimum spanning tree is an optimal Steiner-tree. All the remaining trees of the MSFBCP solution correspond to unused Steiner-points (2-node balanced trees with cost zero), which do not impact the solution cost $(c(P) = w(G') + 0 = w(G'))$. The MSFBCP solution is a feasible and optimal solution for the Steiner-tree problem.

%%%%%%%%%%%%%%%%%%%%%%%%%%%%%%%%%%%%%%%%%%%%%%%%%%%%%%%%%%%%%%%%%%%%%%%%
%%%%%%%%%%%%%%%%%%%%%%%%%%%%%%%%%%%%%%%%%%%%%%%%%%%%%%%%%%%%%%%%%%%%%%

In the second case, since there are disjoint trees containing the terminal vertices, we can assume that there is at least one unused edge of zero cost connecting a pair of vertices that replaces a Steiner point. Indeed, by contradiction, if all the zero-cost edges are used in the solution, then there must be at least one unbalanced tree in the disjoint set of trees. The connection of such a pair of vertices would add an excess of plus or minus one to that tree, making it unbalanced, as illustrated in Figure~\ref{fig:msfbc_infeasible}. Since the trees must be balanced, this solution would be infeasible for the MSFBCP.

We use this observation to generate in polynomial time another solution with the same cost, in which disjoint trees can be merged and at least one more zero-cost edge is included.
Let $T_{1} = (V_{1}, E_{1})$ and $T_{2} = (V_{2}, E_{2})$ be two disjoint balanced trees containing terminal vertices and let $E_{1,2}$ be the set of edges connecting vertices in $T_{1}$ and $T_{2}$. We can use Kruskal's algorithm to compute the minimum spanning tree over the merged component $T_{m}$ = ($V_{m}$, $E_{m}$), where $V_{m}$ = \{$V_{1} \cup V_{2}$\} and $E_{m}$ = \{$E_{1} \cup E_{2} \cup E_{1,2}$\}. The algorithm enumerates the edge set in increasing order, and ties can be broken in such a way that zero-cost edges in $E_{1,2}$ are treated first as well as edges in $T_{1}$ and $T_{2}$. The addition of such edges will connect the disjoint components, and the cost of the new component will be given by
\begin{equation}
c(T_{m}) = c(T_{1}) + c(T_{2}) + 0 = c(T_{1}) + c(T_{2}).
\end{equation}

This ``merge'' procedure can be executed iteratively until a single tree spans all the terminal vertices, resulting in the first case where $c(P) = w(G')$. Since we are able to build a valid and optimal Steiner-tree by solving the MSFBCP on the reduced instance and performing a limited number of calls to a polynomial ``merge'' algorithm, we prove by extension that the MSFBCP is also $\mathcal{NP}$-hard.

\section*{Comparison of linear relaxations}

Let $Z_\textsc{dir}$ and  $Z_\textsc{undir}$ be the linear relaxations of the directed and undirected MSFBCP cut formulations, respectively. We will show that $Z_\textsc{dir} \geq Z_\textsc{undir}$, and that this inequality is strict for some instances.
The undirected formulation of the MSFBCP is formulated in Equations~\mbox{(\ref{c02b})--(\ref{c22b})}. A binary variable $\bar{x}_{ij}$ is associated to each edge $\{i,j\} \in E$, and for each cut $S \subset V$, $\delta(S) = \{ \{i,j\} \in E: |\{i,j\} \cap S| = 1\}$ is the set of edges with exactly one endpoint in $S$.
\begin{align}
\min &\sum_{\{i,j\} \in E} d_{ij} \bar{x}_{ij} \label{c02b} \\
\text{s.t.}  &\sum_{ \{i,j\} \in \delta(S) }  \bar{x}_{ij} \geq 1 & \forall S \subset V: w(S) \neq 0  \label{c32b}\\
                  & \bar{x}_{ij} \in \{0, 1 \}                                     & \forall \{i,j\} \in E  \label{c22b}
\end{align}

\begin{figure}[htbp]
\centering
\includegraphics[scale=1.3]{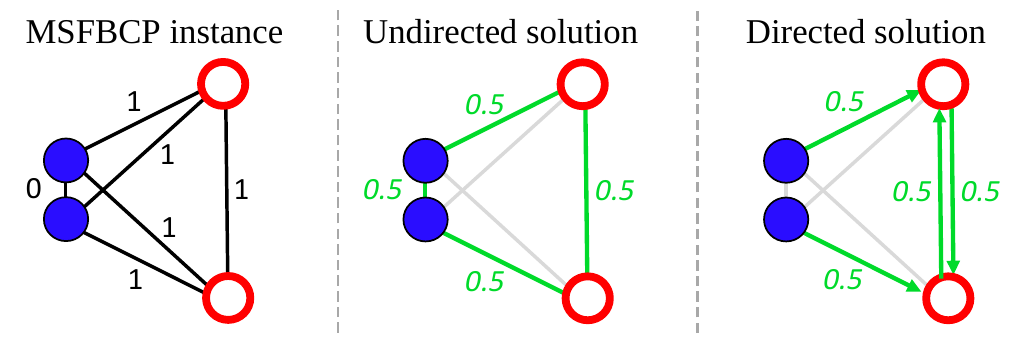}
\caption{Example instance for which $Z_\textsc{dir} > Z_\textsc{undir}$}
\label{fig:LPrelax}
\end{figure}

First, observe that for every feasible solution $\mathbf{x}$ of the directed formulation, the solution~$\mathbf{\bar{x}}$ defined as $\bar{x}_{ij} = x_{ij} + x_{ji}$ for $\{i,j\} \in E$ satisfies Equations \mbox{(\ref{c02b})--(\ref{c22b})}, and has identical cost. As such, $Z_\textsc{dir} \geq Z_\textsc{undir}$. Second, to highlight an instance where $Z_\textsc{dir} > Z_\textsc{undir}$, we adapt a classic example in the literature, illustrated in left side of Figure \ref{fig:LPrelax}. For this instance, the linear relaxation of the undirected formulation has value $1.5$, and that of the directed formulation has value $2$.

\section*{Detailed computational results}

This section presents the detailed results for individual instances for all the benchmark sets considered in this paper. To evaluate the methods, we collected all the known optimal solutions and the best solutions found throughout our tests. These values are presented in the column \textbf{BKS}. We use asterisks to indicate optimal solutions. The best known primal solutions are highlighted in bold.\\

Tables \ref{tab:hybrid_1} to \ref{tab:hybrid_3} present the HILS results for each instance. The value \textbf{Z$_\textsc{Best}$} is the best solution found in 10 runs, while \textbf{Z$_\textsc{Avg}$} is the average solution over these runs. The columns \textbf{GAP$_\textsc{best}$ (\%)} and \textbf{GAP$_\textsc{Avg}$ (\%)} give the percentage gap between the solution and the BKS of the best and average solutions of the 10 runs. Finally, the column \textbf{T(s)} gives the average CPU time per run.\\

Tables \ref{tab:bc} to \ref{tab:bc5} present the branch-and-cut results for every instance. The time limit is set to 3600 seconds. The following information is reported:

\begin{itemize}
\item \textbf{LP$_\textsc{ROOT}$}: the LP value at the root node;
\item \textbf{GAP$_\textsc{ROOT}$}: the percentage gap between the root node LP and the best upper bound
\item \textbf{LB} and \textbf{UB}: the best lower and upper bounds obtained;
\item \textbf{GAP$_\textsc{FINAL}$}: the percentage gap between the best upper and lower bounds;
\item \textbf{T(s)}: the CPU time measured in seconds;
\item {\boldmath $T_{FLOW}(s)$}: the total CPU time spent in the separation algorithm;
\item {\boldmath $T_{ROOT}(s)$}: the total CPU time spent solving the root node;
\item \textbf{N}: the number of nodes explored in the branch-and-bound tree;
\item {\boldmath $N_{TREE}$}: the number of trees in the optimal solution (when available).
\item {\boldmath $S_{TREES}$}: the average size of the trees in the optimal solution (when available).
\end{itemize}

\section*{Benchmark Instances}
The benchmark instances used in this study can be found in the online supplement or at the following address: \url{https://w1.cirrelt.ca/~vidalt/en/research-data.html}.

\newpage

\begin{table}[htbp]
\renewcommand{\arraystretch}{1.4}
\centering
\setlength{\tabcolsep}{.7em}
\caption{Detailed results for HILS algorithm}
\label{tab:hybrid_1}
\scalebox{0.82}
{
\begin{tabular}{|c|c|c|c|c|c|c|c|c|}
\hline
Instance & $\abs{V}$& $\abs{E}$ & Z$_\textsc{Best}$ & GAP$_\textsc{Best}$(\%) & Z$_\textsc{Avg}$ & GAP$_\textsc{avg}$(\%) & BKS  & Avg T(s) \\ \hline
PUC-8-1 & 10 & 90 & \textbf{24.71} & 0.00 & \textbf{24.71} & 0.00 & 24.71* & 0.40 \\
PUC-8-2 & 10 & 90 & \textbf{16.46} & 0.00 & \textbf{16.46} & 0.00 & 16.46* & 0.25 \\
PUC-8-3 & 10 & 90 & \textbf{14.26} & 0.00 & \textbf{14.26} & 0.00 & 14.26* & 0.30 \\
PUC-8-4 & 10 & 90 & \textbf{24.99} & 0.00 & \textbf{24.99} & 0.00 & 24.99* & 0.23 \\
PUC-8-5 & 10 & 90 & \textbf{23.90} & 0.00 & 27.29 & 12.42 & 23.90* & 0.15 \\
PUC-12-1 & 14 & 182 & \textbf{54.54} & 0.00 & \textbf{54.54} & 0.00 & 54.54* & 0.86 \\
PUC-12-2 & 14 & 182 & \textbf{49.05} & 0.00 & \textbf{49.05} & 0.00 & 49.05* & 0.67 \\
PUC-12-3 & 14 & 182 & \textbf{53.59} & 0.00 & \textbf{53.59} & 0.00 & 53.59* & 0.72 \\ 
PUC-12-4 & 14 & 182 & \textbf{49.49} & 0.00 & \textbf{49.49} & 0.00 & 49.49* & 0.94 \\
PUC-12-5 & 14 & 182 & \textbf{46.94} & 0.00 & \textbf{46.94} & 0.00 & 46.94* & 0.66 \\
PUC-16-1 & 18 & 306 & \textbf{79.38} & 0.00 & \textbf{79.38} & 0.00 & 79.38* & 1.46 \\
PUC-16-2 & 18 & 306 & \textbf{69.68} & 0.00 & \textbf{69.68} & 0.00 & 69.68* & 1.46 \\
PUC-16-3 & 18 & 306 & \textbf{82.50} & 0.00 & 82.86 & 0.43 & 82.50* & 1.57 \\
PUC-16-4 & 18 & 306 & \textbf{79.01} & 0.00 & \textbf{79.01} & 0.00 & 79.01* & 1.46 \\
PUC-16-5 & 18 & 306 & \textbf{79.60} & 0.00 & 81.38 & 2.19 & 79.60* & 2.05 \\
PUC-20-1 & 22 & 462 & \textbf{128.16} & 0.00 & 128.20 & 0.03 & 128.16* & 2.15 \\
PUC-20-2 & 22 & 462 & \textbf{132.93} & 0.00 & \textbf{132.93} & 0.00 & 132.93* & 3.73 \\
PUC-20-3 & 22 & 462 & \textbf{112.17} & 0.00 & \textbf{112.17} & 0.00 & 112.17* & 3.59 \\
PUC-20-4 & 22 & 462 & \textbf{119.74} & 0.00 & 120.97 & 1.02 & 119.74* & 3.88 \\
PUC-20-5 & 22 & 462 & \textbf{121.93} & 0.00 & \textbf{121.93} & 0.00 & 121.93* & 4.64 \\ 
PUC-24-1 & 26 & 650 & \textbf{183.25} & 0.00 & 183.97 & 0.39 & 183.25* & 4.15 \\
PUC-24-2 & 26 & 650 & \textbf{124.07} & 0.00 & 125.96 & 1.50 & 124.07* & 6.30 \\ 
PUC-24-3 & 26 & 650 & \textbf{157.61} & 0.00 & 157.64 & 0.02 & 157.61* & 4.81 \\
PUC-24-4 & 26 & 650 & \textbf{163.07} & 0.00 & 163.50 & 0.26 & 163.07* & 5.67 \\
PUC-24-5 & 26 & 650 & \textbf{168.41} & 0.00 & 170.16 & 1.03 & 168.41* & 6.36 \\
PUC-28-1 & 30 & 870 & \textbf{203.86} & 0.00 & 208.37 & 2.16 & 203.86* & 8.87 \\
PUC-28-2 & 30 & 870 & \textbf{209.82} & 0.00 & 215.03 & 2.42 & 209.82* & 12.87 \\
PUC-28-3 & 30 & 870 & \textbf{215.69} & 0.00 & \textbf{215.69} & 0.00 & 215.69* & 7.89 \\
PUC-28-4 & 30 & 870 & \textbf{207.88} & 0.00 & 208.13 & 0.12 & 207.88* & 10.82 \\
PUC-28-5 & 30 & 870 & \textbf{209.48} & 0.00 & 210.69 & 0.57 & 209.48* & 12.04 \\
PUC-32-1 & 34 & 1122 & \textbf{259.67} & 0.00 & \textbf{259.67} & 0.00 & 259.67* & 12.33 \\
PUC-32-2 & 34 & 1122 & \textbf{237.03} & 0.00 & 237.26 & 0.10 & 237.03* & 16.99 \\  
PUC-32-3 & 34 & 1122 & \textbf{222.04} & 0.00 & 222.56 & 0.23 & 222.04* & 16.07 \\
PUC-32-4 & 34 & 1122 & \textbf{234.29} & 0.00 & 237.36 & 1.29 & 234.29* & 13.36 \\
PUC-32-5 & 34 & 1122 & \textbf{284.18} & 0.00 & 290.45 & 2.16 & 284.18* & 15.66 \\
\hline
\end{tabular}
}
\end{table}
\begin{table}[htbp]
\renewcommand{\arraystretch}{1.4}
\centering
\setlength{\tabcolsep}{.7em}
\caption{Detailed results for HILS algorithm (continued)}
\label{tab:hybrid_2}
\scalebox{0.82}
{
\begin{tabular}{|c|c|c|c|c|c|c|c|c|}
\hline
Instance & $\abs{V}$ & $\abs{E}$ & Z$_\textsc{Best}$ & GAP$_\textsc{Best}$(\%) & Z$_\textsc{Avg}$ & GAP$_\textsc{avg}$(\%) & BKS  & Avg T(s) \\ \hline
PUC-36-1 & 38 & 1406 & \textbf{297.53} & 0.00 & 297.70 & 0.06 & 297.53* & 18.39 \\
PUC-36-2 & 38 & 1406 & \textbf{302.64} & 0.00 & 303.97 & 0.44 & 302.64* & 21.36 \\
PUC-36-3 & 38 & 1406 & \textbf{291.27} & 0.00 & 292.30 & 0.35 & 291.27* & 21.21 \\
PUC-36-4 & 38 & 1406 & \textbf{286.08} & 0.00 & 296.15 & 3.40 & 286.08* & 16.51 \\
PUC-36-5 & 38 & 1406 & \textbf{263.76} & 0.00 & 264.96 & 0.45 & 263.76* & 22.10 \\
PUC-40-1 & 42 & 1722 & \textbf{347.49} & 0.00 & 351.70 & 1.20 & 347.49* & 48.26 \\
PUC-40-2 & 42 & 1722 & \textbf{381.25} & 0.00 & \textbf{381.25} & 0.00 & 381.25* & 31.67 \\
PUC-40-3 & 42 & 1722 & \textbf{391.30} & 0.00 & 394.76 & 0.88 & 391.30* & 24.12 \\
PUC-40-4 & 42 & 1722 & \textbf{399.45} & 0.00 & 402.28 & 0.70 & 399.45* & 24.86 \\
PUC-40-5 & 42 & 1722 & \textbf{366.81} & 0.00 & 368.18 & 0.37 & 366.81* & 33.52 \\
PUC-44-1 & 46 & 2070 & 462.14 & 0.68 & 473.89 & 3.14 & 459.01* & 53.56 \\
PUC-44-2 & 46 & 2070 & \textbf{413.55} & 0.00 & 417.59 & 0.97 & 413.55* & 36.88 \\
PUC-44-3 & 46 & 2070 & \textbf{433.65} & 0.00 & 439.33 & 1.29 & 434.67* & 54.80 \\
PUC-44-4 & 46 & 2070 & \textbf{461.95} & 0.00 & 464.01 & 0.44 & 461.95* & 33.06 \\
PUC-44-5 & 46 & 2070 & \textbf{448.58} & 0.00 & 454.66 & 1.34 & 448.58* & 44.20 \\
PUC-48-1 & 50 & 2450 & \textbf{460.30} & 0.00 & 464.82 & 0.97 & 460.75* & 57.49 \\
PUC-48-2 & 50 & 2450 & \textbf{477.41} & 0.00 & 480.90 & 0.73 & 477.41* & 40.90 \\
PUC-48-3 & 50 & 2450 & \textbf{418.47} & 0.00 & 421.27 & 0.66 & 418.47* & 53.59 \\
PUC-48-4 & 50 & 2450 & \textbf{459.07} & 0.00 & 463.97 & 1.06 & 459.07* & 48.02 \\
PUC-48-5 & 50 & 2450 & \textbf{480.62} & 0.00 & 481.82 & 0.25 & 480.62* & 44.56 \\
PUC-52-1 & 54 & 2862 & \textbf{574.00} & 0.00 & 584.11 & 1.73 & 574.00* & 81.22 \\
PUC-52-2 & 54 & 2862 & \textbf{515.26} & 0.00 & 526.52 & 2.14 & 515.26* & 84.54 \\
PUC-52-3 & 54 & 2862 & \textbf{594.35} & 0.00 & 595.50 & 0.19 & 594.35* & 56.76 \\
PUC-52-4 & 54 & 2862 & \textbf{472.59} & 0.00 & 475.03 & 0.51 & 472.59* & 58.93 \\
PUC-52-5 & 54 & 2862 & \textbf{585.80} & 0.00 & 598.28 & 2.09 & 585.80* & 68.72 \\
PUC-56-1 & 58 & 3306 & \textbf{640.93} & 0.00 & 657.44 & 2.51 & 640.93* & 91.59 \\
PUC-56-2 & 58 & 3306 & \textbf{631.74} & 0.00 & 636.72 & 0.78 & 631.74* & 82.25 \\
PUC-56-3 & 58 & 3306 & \textbf{594.23} & 0.00 & 594.43 & 0.03 & 594.23* & 73.93 \\
PUC-56-4 & 58 & 3306 & \textbf{570.74} & 0.00 & 576.15 & 0.94 & 570.74* & 91.29 \\
PUC-56-5 & 58 & 3306 & \textbf{544.91} & 0.00 & 558.62 & 2.45 & 544.91* & 74.74 \\
PUC-60-1 & 62 & 3782 & \textbf{639.40} & 0.00 & 648.90 & 1.46 & 639.40* & 93.92 \\
PUC-60-2 & 62 & 3782 & \textbf{722.47} & 0.00 & 736.34 & 1.88 & 722.47* & 98.70 \\
PUC-60-3 & 62 & 3782 & \textbf{732.90} & 0.00 & 746.47 & 1.82 & 732.90* & 119.74 \\
PUC-60-4 & 62 & 3782 & \textbf{668.18} & 0.00 & 669.94 & 0.26 & 668.18* & 92.07 \\
PUC-60-5 & 62 & 3782 & \textbf{706.65} & 0.00 & 709.07 & 0.34 & 706.65* & 81.20 \\ 
PUC-64-1 & 66 & 4290 & 802.77 & 2.03 & 834.17 & 5.68 & 786.81* & 145.21 \\
\hline
\end{tabular}
}
\end{table}
\begin{table}[htbp]
\renewcommand{\arraystretch}{1.4}
\setlength{\tabcolsep}{.7em}
\caption{Detailed results for HILS algorithm (continued)}
\label{tab:hybrid_3}
\hspace*{-0.5cm}
\scalebox{0.82}
{
\begin{tabular}{|c|c|c|c|c|c|c|c|c|}
\hline
Instance & $\abs{V}$ & $\abs{E}$ & Z$_\textsc{Best}$ & GAP$_\textsc{Best}$(\%) & Z$_\textsc{Avg}$ & GAP$_\textsc{avg}$(\%) & BKS  & Avg T(s) \\ \hline
PUC-64-2 & 66 & 4290 & \textbf{748.66} & 0.00 & 797.03 & 6.07 & 748.66* & 126.61 \\
PUC-64-3 & 66 & 4290 & \textbf{754.85} & 0.00 & 757.45 & 0.34 & 754.85* & 131.41 \\
PUC-64-4 & 66 & 4290 & \textbf{805.77} & 0.00 & 825.49 & 2.39 & 805.77* & 141.08 \\
PUC-64-5 & 66 & 4290 & \textbf{789.85} & 0.00 & 796.73 & 0.86 & 789.85* & 127.95 \\
PUC-80-1 & 82 & 6642 & 1089.77 & 1.20 & 1102.81 & 2.36 & 1076.83* & 280.25 \\
PUC-80-2 & 82 & 6642 & 1084.38 & 0.80 & 1124.80 & 4.36 & 1075.81* & 295.59 \\
PUC-80-3 & 82 & 6642 & \textbf{1048.91} & 0.00 & 1079.11 & 2.80 & 1048.91* & 266.89 \\
PUC-80-4 & 82 & 6642 & \textbf{1114.46} & 0.00 & 1135.72 & 1.87 & 1114.46* & 329.72 \\
PUC-80-5 & 82 & 6642 & \textbf{1122.35} & 0.00 & 1180.16 & 4.90 & 1122.35* & 350.77 \\
PUC-96-1 & 98 & 9506 & \textbf{1348.94} & 0.00 & 1421.62 & 5.11 & 1348.94* & 643.02 \\
PUC-96-2 & 98 & 9506 & 1613.75 & 1.22 & 1698.21 & 6.12 & 1594.22* & 892.48 \\
PUC-96-3 & 98 & 9506 & 1511.63 & 0.48 & 1569.20 & 4.13 & 1504.42* & 514.21 \\
PUC-96-4 & 98 & 9506 & 1495.14 & 0.51 & 1583.95 & 6.09 & 1487.52* & 706.41 \\
PUC-96-5 & 98 & 9506 & \textbf{1422.95} & 0.00 & 1485.31 & 4.20 & 1422.95* & 494.14 \\
PUC-128-1 & 130 & 16770 & 2450.87 & 1.38 & 2579.99 & 6.30 & 2417.43* & 2482.93 \\
PUC-128-2 & 130 & 16770 & \textbf{2166.76} & 0.00 & 2322.01 & 6.69 & 2166.76* & 1099.27 \\
PUC-128-3 & 130 & 16770 & 2576.42 & 7.56 & 2624.88 & 8.75 & 2395.29* & 3348.55 \\
PUC-128-4 & 130 & 16770 & 2293.61 & 0.10 & 2485.22 & 7.81 & 2291.24* & 1735.70 \\
PUC-128-5 & 130 & 16770 & \textbf{2269.25} & 0.00 & 2482.12 & 8.58 & 2269.25* & 1791.81 \\
PUC-256-1 & 258 & 66306 & \textbf{7061.17} & 0.00 & 7643.75 & 7.62 & 7061.17 & 3600.00 \\
PUC-256-2 & 258 & 66306 & \textbf{7748.26} & 0.00 & 8008.76 & 3.25 & 7748.27 & 3600.00 \\
PUC-256-3 & 258 & 66306 & \textbf{7518.65} & 0.00 & 7824.09 & 3.90 & 7518.65 & 3600.00 \\
PUC-256-4 & 258 & 66306 & \textbf{7530.24} & 0.00 & 7539.17 & 0.12 & 7530.24 & 3600.00 \\
PUC-256-5 & 258 & 66306 & \textbf{6913.37} & 0.00 & 8038.88 & 14.00 & 6913.38 & 3600.00 \\
PUC-512-1 & 514 & 263682 & \textbf{23517.25} & 0.00 & 26140.36 & 10.03 & 23517.26 & 3600.00 \\
PUC-512-2 & 514 & 263682 & \textbf{23286.79} & 0.00 & 23978.25 & 2.88 & 23286.79 & 3600.00 \\
PUC-512-3 & 514 & 263682 & \textbf{22874.35} & 0.00 & 25420.15 & 10.01 & 22874.35 & 3600.00 \\
PUC-512-4 & 514 & 263682 & \textbf{23442.23} & 0.00 & 24726.64 & 5.19 & 23442.23 & 3600.00 \\
PUC-512-5 & 514 & 263682 & \textbf{22863.50} & 0.00 & 23672.32 & 3.42 & 22863.50 & 3600.00 \\
PUC-1024-1 & 1026 & 1051650 & \textbf{71011.17} & 0.00 & 72375.26 & 1.88 & 71011.17 & 3600.00 \\
PUC-1024-2 & 1026 & 1051650 & \textbf{69510.15} & 0.00 & 69890.54 & 0.54 & 69510.15 & 3600.00 \\
PUC-1024-3 & 1026 & 1051650 & \textbf{69401.36} & 0.00 & 69523.12 & 0.18 & 69401.36 & 3600.00 \\
PUC-1024-4 & 1026 & 1051650 & \textbf{68040.88} & 0.00 & 69141.38 & 1.59 & 68040.88 & 3600.00 \\
PUC-1024-5 & 1026 & 1051650 & \textbf{69884.32} & 0.00 & 71384.11 & 2.10 & 69884.32 & 3600.00  \\ \hline
\end{tabular}
}
\end{table}

{
\begin{landscape}
\begin{table}[htbp]
\renewcommand{\arraystretch}{1.4}
\centering
\hspace{-1.1cm}
\caption{Detailed results for B\&C algorithm}
\label{tab:bc}
\hspace{-1.1cm}
\scalebox{0.95}{
\begin{tabular}{|c|ccccc|ccc|ccc|}
\hline
Group & LP$_\textsc{Root}$ & GAP$_\textsc{Root}$(\%) & LB & UB & GAP$_\textsc{Final}$(\%) & T(s) & T$_\textsc{Flow}$(s) & T$_\textsc{Root}$(s) & N$_\textsc{Node}$ & N$_\textsc{Tree}$ & S$_\textsc{Tree}$ \\ \hline 
PUC-8-1 & 24.71 & 0.00 & 24.71 & 24.71 & 0.00 & $<$0.01 & $<$0.01 & $<$0.01 & 1 & 2 & 5.00\\
PUC-8-2 & 16.46 & 0.00 & 16.46 & 16.46 & 0.00 & $<$0.01 & $<$0.01 & $<$0.01 & 1 & 3 & 3.33\\
PUC-8-3 & 14.26 & 0.00 & 14.26 & 14.26 & 0.00 & $<$0.01 & $<$0.01 & $<$0.01 & 1 & 2 & 5.00\\
PUC-8-4 & 24.99 & 0.00 & 24.99 & 24.99 & 0.00 & $<$0.01 & $<$0.01 & $<$0.01 & 1 & 2 & 5.00\\
PUC-8-5 & 23.90 & 0.00 & 23.90 & 23.90 & 0.00 & $<$0.01 & $<$0.01 & $<$0.01 & 1 & 2 & 5.00\\
PUC-12-1 & 54.54 & 0.00 & 54.54 & 54.54 & 0.00 & $<$0.01 & $<$0.01 & $<$0.01 & 1 & 3 & 4.67\\
PUC-12-2 & 49.05 & 0.00 & 49.05 & 49.05 & 0.00 & $<$0.01 & $<$0.01 & $<$0.01 & 1 & 5 & 2.80\\
PUC-12-3 & 53.59 & 0.00 & 53.59 & 53.59 & 0.00 & $<$0.01 & $<$0.01 & $<$0.01 & 1 & 4 & 3.50\\
PUC-12-4 & 49.49 & 0.00 & 49.49 & 49.49 & 0.00 & $<$0.01 & $<$0.01 & $<$0.01 & 1 & 3 & 4.67\\
PUC-12-5 & 46.94 & 0.00 & 46.94 & 46.94 & 0.00 & $<$0.01 & $<$0.01 & $<$0.01 & 1 & 3 & 4.67\\
PUC-16-1 & 79.38 & 0.00 & 79.38 & 79.38 & 0.00 & $<$0.01 & $<$0.01 & $<$0.01 & 1 & 4 & 4.50\\
PUC-16-2 & 69.68 & 0.00 & 69.68 & 69.68 & 0.00 & $<$0.01 & $<$0.01 & $<$0.01 & 1 & 4 & 4.50\\
PUC-16-3 & 82.50 & 0.00 & 82.50 & 82.50 & 0.00 & $<$0.01 & $<$0.01 & $<$0.01 & 1 & 5 & 3.60\\
PUC-16-4 & 79.01 & 0.00 & 79.01 & 79.01 & 0.00 & $<$0.01 & $<$0.01 & $<$0.01 & 1 & 8 & 2.25\\
PUC-16-5 & 79.60 & 0.00 & 79.60 & 79.60 & 0.00 & $<$0.01 & $<$0.01 & $<$0.01 & 1 & 3 & 6.00\\
PUC-20-1 & 128.16 & 0.00 & 128.16 & 128.16 & 0.00 & $<$0.01 & $<$0.01 & $<$0.01 & 1 & 5 & 4.40\\
PUC-20-2 & 132.93 & 0.00 & 132.93 & 132.93 & 0.00 & $<$0.01 & $<$0.01 & $<$0.01 & 1 & 6 & 3.67\\
PUC-20-3 & 112.17 & 0.00 & 112.17 & 112.17 & 0.00 & $<$0.01 & $<$0.01 & $<$0.01 & 1 & 8 & 2.75\\
PUC-20-4 & 114.71 & 4.20 & 119.74 & 119.74 & 0.00 & $<$0.01 & $<$0.01 & $<$0.01 & 1 & 7 & 3.14\\
PUC-20-5 & 119.73 & 1.81 & 121.93 & 121.93 & 0.00 & $<$0.01 & $<$0.01 & $<$0.01 & 1 & 4 & 5.50\\
PUC-24-1 & 181.39 & 1.01 & 183.25 & 183.25 & 0.00 & $<$0.01 & $<$0.01 & $<$0.01 & 1 & 6 & 4.33 \\ \hline
\end{tabular}
}
\end{table}
\end{landscape}
}
{
\begin{landscape}
\begin{table}[htbp]
\renewcommand{\arraystretch}{1.4}
\centering
\hspace{-1.1cm}
\caption{Detailed results for B\&C algorithm (continued)}
\label{tab:bc2}
\hspace{-1.1cm}
\scalebox{0.91}{
\begin{tabular}{|c|ccccc|ccc|ccc|}
\hline
Group & LP$_\textsc{Root}$ & GAP$_\textsc{Root}$(\%) & LB & UB & GAP$_\textsc{Final}$(\%) & T(s) & T$_\textsc{Flow}$(s) & T$_\textsc{Root}$(s) & N$_\textsc{Node}$ & N$_\textsc{Tree}$ & S$_\textsc{Tree}$ \\ \hline PUC-24-2 & 124.07 & 0.00 & 124.07 & 124.07 & 0.00 & $<$0.01 & $<$0.01 & $<$0.01 & 1 & 9 & 2.89\\
PUC-24-3 & 155.15 & 1.56 & 157.61 & 157.61 & 0.00 & $<$0.01 & $<$0.01 & $<$0.01 & 1 & 9 & 2.89\\
PUC-24-4 & 160.97 & 1.28 & 163.07 & 163.07 & 0.00 & $<$0.01 & $<$0.01 & $<$0.01 & 1 & 6 & 4.33\\
PUC-24-5 & 164.09 & 2.57 & 168.41 & 168.41 & 0.00 & $<$0.01 & $<$0.01 & $<$0.01 & 1 & 7 & 3.71\\
PUC-28-1 & 194.91 & 4.39 & 203.86 & 203.86 & 0.00 & $<$0.01 & $<$0.01 & $<$0.01 & 1 & 10 & 3.00\\
PUC-28-2 & 201.79 & 3.83 & 209.82 & 209.82 & 0.00 & $<$0.01 & $<$0.01 & $<$0.01 & 1 & 8 & 3.75\\
PUC-28-3 & 213.65 & 0.94 & 215.69 & 215.69 & 0.00 & $<$0.01 & $<$0.01 & $<$0.01 & 1 & 6 & 5.00\\
PUC-28-4 & 207.88 & 0.00 & 207.88 & 207.88 & 0.00 & $<$0.01 & $<$0.01 & $<$0.01 & 1 & 7 & 4.29\\
PUC-28-5 & 208.35 & 0.54 & 209.48 & 209.48 & 0.00 & $<$0.01 & $<$0.01 & $<$0.01 & 1 & 7 & 4.29\\
PUC-32-1 & 259.67 & 0.00 & 259.67 & 259.67 & 0.00 & $<$0.01 & $<$0.01 & $<$0.01 & 1 & 10 & 3.40\\
PUC-32-2 & 233.93 & 1.31 & 237.03 & 237.03 & 0.00 & $<$0.01 & $<$0.01 & $<$0.01 & 1 & 9 & 3.78\\
PUC-32-3 & 222.04 & 0.00 & 222.04 & 222.04 & 0.00 & $<$0.01 & $<$0.01 & $<$0.01 & 1 & 12 & 2.83\\
PUC-32-4 & 226.04 & 3.52 & 234.29 & 234.29 & 0.00 & $<$0.01 & $<$0.01 & $<$0.01 & 1 & 13 & 2.62\\
PUC-32-5 & 261.18 & 8.09 & 284.18 & 284.18 & 0.00 & 0.01 & $<$0.01 & 0.01 & 1 & 9 & 3.78\\
PUC-36-1 & 280.47 & 5.74 & 297.53 & 297.53 & 0.00 & $<$0.01 & $<$0.01 & $<$0.01 & 1 & 9 & 4.22\\
PUC-36-2 & 296.88 & 1.90 & 302.64 & 302.64 & 0.00 & $<$0.01 & $<$0.01 & $<$0.01 & 1 & 10 & 3.80\\
PUC-36-3 & 282.69 & 2.94 & 291.27 & 291.27 & 0.00 & $<$0.01 & $<$0.01 & $<$0.01 & 1 & 9 & 4.22\\
PUC-36-4 & 268.05 & 6.30 & 286.08 & 286.08 & 0.00 & $<$0.01 & $<$0.01 & $<$0.01 & 1 & 13 & 2.92\\
PUC-36-5 & 257.21 & 2.49 & 263.76 & 263.76 & 0.00 & $<$0.01 & $<$0.01 & $<$0.01 & 1 & 9 & 4.22\\
PUC-40-1 & 335.32 & 3.50 & 347.49 & 347.49 & 0.00 & $<$0.01 & $<$0.01 & $<$0.01 & 1 & 11 & 3.82\\
PUC-40-2 & 369.86 & 2.99 & 381.25 & 381.25 & 0.00 & $<$0.01 & $<$0.01 & $<$0.01 & 1 & 11 & 3.82\\
PUC-40-3 & 370.06 & 5.43 & 391.30 & 391.30 & 0.00 & $<$0.01 & $<$0.01 & $<$0.01 & 1 & 11 & 3.82\\
PUC-40-4 & 394.30 & 1.29 & 399.45 & 399.45 & 0.00 & 2.34 & 0.19 & 0.05 & 55 & 14 & 3.00 \\ \hline
\end{tabular}
}
\end{table}
\end{landscape}
}
{
\begin{landscape}
\begin{table}[htbp]
\renewcommand{\arraystretch}{1.4}
\centering
\hspace{-1.2cm}
\caption{Detailed results for B\&C algorithm (continued)}
\label{tab:bc3}
\hspace{-1.5cm}
\scalebox{0.90}{
\begin{tabular}{|c|ccccc|ccc|ccc|}
\hline
Group & LP$_\textsc{Root}$ & GAP$_\textsc{Root}$(\%) & LB & UB & GAP$_\textsc{Final}$(\%) & T(s) & T$_\textsc{Flow}$(s) & T$_\textsc{Root}$(s) & N$_\textsc{Node}$ & N$_\textsc{Tree}$ & S$_\textsc{Tree}$ \\ \hline 
PUC-40-5 & 356.37 & 2.85 & 366.81 & 366.81 & 0.00 & $<$0.01 & $<$0.01 & $<$0.01 & 1 & 16 & 2.63\\
PUC-44-1 & 451.81 & 1.57 & 459.01 & 459.01 & 0.00 & 9.52 & 0.75 & 0.06 & 113 & 10 & 4.60\\
PUC-44-2 & 405.29 & 2.00 & 413.55 & 413.55 & 0.00 & $<$0.01 & $<$0.01 & $<$0.01 & 1 & 17 & 2.71\\
PUC-44-3 & 404.95 & 6.62 & 433.65 & 433.65 & 0.00 & $<$0.01 & $<$0.01 & $<$0.01 & 1 & 13 & 3.54\\
PUC-44-4 & 446.29 & 3.39 & 461.95 & 461.95 & 0.00 & $<$0.01 & $<$0.01 & $<$0.01 & 1 & 15 & 3.07\\
PUC-44-5 & 423.41 & 5.61 & 448.58 & 448.58 & 0.00 & 0.05 & $<$0.01 & 0.05 & 1 & 14 & 3.29\\
PUC-48-1 & 437.59 & 4.93 & 460.30 & 460.30 & 0.00 & $<$0.01 & $<$0.01 & $<$0.01 & 1 & 13 & 3.85\\
PUC-48-2 & 460.04 & 3.64 & 477.41 & 477.41 & 0.00 & $<$0.01 & $<$0.01 & $<$0.01 & 1 & 19 & 2.63\\
PUC-48-3 & 411.55 & 1.65 & 418.47 & 418.47 & 0.00 & $<$0.01 & $<$0.01 & $<$0.01 & 1 & 19 & 2.63\\
PUC-48-4 & 447.94 & 2.42 & 459.07 & 459.07 & 0.00 & $<$0.01 & $<$0.01 & $<$0.01 & 1 & 16 & 3.13\\
PUC-48-5 & 470.43 & 2.12 & 480.62 & 480.62 & 0.00 & $<$0.01 & $<$0.01 & $<$0.01 & 1 & 11 & 4.55\\
PUC-52-1 & 542.87 & 5.42 & 574.00 & 574.00 & 0.00 & $<$0.01 & $<$0.01 & $<$0.01 & 1 & 20 & 2.70\\
PUC-52-2 & 501.23 & 2.72 & 515.26 & 515.26 & 0.00 & $<$0.01 & $<$0.01 & $<$0.01 & 1 & 19 & 2.84\\
PUC-52-3 & 574.04 & 3.42 & 594.35 & 594.35 & 0.00 & $<$0.01 & $<$0.01 & $<$0.01 & 1 & 17 & 3.18\\
PUC-52-4 & 447.91 & 5.22 & 472.59 & 472.59 & 0.00 & $<$0.01 & $<$0.01 & $<$0.01 & 1 & 21 & 2.57\\
PUC-52-5 & 551.32 & 5.88 & 585.80 & 585.80 & 0.00 & $<$0.01 & $<$0.01 & $<$0.01 & 1 & 17 & 3.17\\
PUC-56-1 & 605.68 & 5.50 & 640.93 & 640.93 & 0.00 & 0.02 & $<$0.01 & 0.02 & 1 & 12 & 4.83\\
PUC-56-2 & 599.16 & 5.16 & 631.74 & 631.74 & 0.00 & $<$0.01 & $<$0.01 & $<$0.01 & 1 & 18 & 3.22\\
PUC-56-3 & 566.63 & 4.64 & 594.23 & 594.23 & 0.00 & $<$0.01 & $<$0.01 & $<$0.01 & 1 & 14 & 4.14\\
PUC-56-4 & 553.13 & 3.09 & 570.74 & 570.74 & 0.00 & $<$0.01 & $<$0.01 & $<$0.01 & 1 & 22 & 2.64\\
PUC-56-5 & 530.04 & 2.73 & 544.91 & 544.91 & 0.00 & $<$0.01 & $<$0.01 & $<$0.01 & 1 & 20 & 2.90\\
PUC-60-1 & 617.71 & 3.39 & 639.40 & 639.40 & 0.00 & 0.01 & $<$0.01 & 0.01 & 1 & 20 & 3.10\\
PUC-60-2 & 717.76 & 0.65 & 722.47 & 722.47 & 0.00 & 1.07 & 0.09 & 0.19 & 9 & 17 & 3.65 \\
\hline
\end{tabular}
}
\end{table}
\end{landscape}
}
{
\begin{landscape}
\begin{table}[htbp]
\renewcommand{\arraystretch}{1.4}
\centering
\hspace{-1.2cm}
\caption{Detailed results for B\&C algorithm (continued)}
\label{tab:bc4}
\hspace{-1.5cm}
\scalebox{0.90}{
\begin{tabular}{|c|ccccc|ccc|ccc|}
\hline
Group & LP$_\textsc{Root}$ & GAP$_\textsc{Root}$(\%) & LB & UB & GAP$_\textsc{Final}$(\%) & T(s) & T$_\textsc{Flow}$(s) & T$_\textsc{Root}$(s) & N$_\textsc{Node}$ & N$_\textsc{Tree}$ & S$_\textsc{Tree}$ \\ \hline 
PUC-60-3 & 732.62 & 0.04 & 732.90 & 732.90 & 0.00 & 0.17 & 0.01 & 0.09 & 3 & 22 & 2.82\\
PUC-60-4 & 642.54 & 3.84 & 668.18 & 668.18 & 0.00 & $<$0.01 & $<$0.01 & $<$0.01 & 1 & 20 & 3.10\\
PUC-60-5 & 671.32 & 5.00 & 706.65 & 706.65 & 0.00 & 0.01 & $<$0.01 & 0.01 & 1 & 19 & 3.26\\
PUC-64-1 & 786.81 & 0.00 & 786.81 & 786.81 & 0.00 & 2.68 & 0.11 & 0.40 & 13 & 17 & 3.88\\
PUC-64-2 & 712.84 & 4.78 & 748.66 & 748.66 & 0.00 & 0.02 & $<$0.01 & 0.02 & 1 & 10 & 19.00\\
PUC-64-3 & 715.25 & 5.25 & 754.85 & 754.85 & 0.00 & 0.01 & $<$0.01 & 0.01 & 1 & 9 & 18.00\\
PUC-64-4 & 798.87 & 0.86 & 805.77 & 805.77 & 0.00 & 12.75 & 0.46 & 0.33 & 33 & 18 & 3.67\\
PUC-64-5 & 786.26 & 0.45 & 789.85 & 789.85 & 0.00 & 0.61 & 0.05 & 0.14 & 3 & 10 & 25.00\\
PUC-80-1 & 1063.93 & 1.20 & 1076.83 & 1076.83 & 0.00 & 46.73 & 3.74 & 0.20 & 149 & 29 & 2.83\\
PUC-80-2 & 1049.38 & 2.46 & 1075.81 & 1075.81 & 0.00 & 405.18 & 37.57 & 0.21 & 1397 & 29 & 2.83\\
PUC-80-3 & 1043.09 & 0.56 & 1048.91 & 1048.91 & 0.00 & 2.13 & 0.25 & 0.22 & 9 & 25 & 3.28\\
PUC-80-4 & 1076.07 & 3.44 & 1114.46 & 1114.46 & 0.00 & 0.01 & 0.22 & 0.01 & 1 & 11 & 26.00\\
PUC-80-5 & 1110.42 & 1.06 & 1122.35 & 1122.35 & 0.00 & 11.22 & 0.79 & 0.41 & 43 & 29 & 2.83\\
PUC-96-1 & 1340.25 & 0.64 & 1348.94 & 1348.94 & 0.00 & 11.88 & 1.27 & 0.41 & 25 & 31 & 3.16\\
PUC-96-2 & 1566.29 & 1.75 & 1594.22 & 1594.22 & 0.00 & 1630.32 & 153.32 & 1.07 & 2251 & 29 & 3.38\\
PUC-96-3 & 1491.40 & 0.86 & 1504.42 & 1504.42 & 0.00 & 615.45 & 36.91 & 1.79 & 685 & 26 & 3.77\\
PUC-96-4 & 1467.85 & 1.32 & 1487.52 & 1487.52 & 0.00 & 49.71 & 5.48 & 0.71 & 103 & 32 & 3.06\\
PUC-96-5 & 1422.95 & 0.00 & 1422.95 & 1422.95 & 0.00 & 0.20 & 0.02 & 0.20 & 1 & 7 & 30.00\\
PUC-128-1 & 2369.39 & 1.99 & 2417.43 & 2417.43 & 0.00 & 3423.57 & 368.07 & 1.52 & 2386 & 38 & 3.42\\
PUC-128-2 & 2156.78 & 0.46 & 2166.77 & 2166.77 & 0.00 & 17.46 & 1.62 & 0.54 & 39 & 44 & 2.96\\
PUC-128-3 & 2374.39 & 0.87 & 2395.29 & 2395.29 & 0.00 & 3388.97 & 328.67 & 1.91 & 3039 & 44 & 2.96\\
PUC-128-4 & 2201.68 & 3.91 & 2291.24 & 2291.24 & 0.00 & 3101.67 & 311.98 & 1.19 & 2162 & 43 & 3.02\\
PUC-128-5 & 2269.26 & 0.00 & 2269.26 & 2269.26 & 0.00 & 0.19 & 0.01 & 0.19 & 1 & 53 & 2.45 \\ \hline
\end{tabular}
}
\end{table}
\end{landscape}
}
{
\begin{landscape}
\begin{table}[h]
\renewcommand{\arraystretch}{1.4}
\vspace{2cm}
\centering
\hspace{-1.2cm}
\caption{Detailed results for B\&C algorithm (continued)}
\label{tab:bc5}
\hspace{-1.5cm}
\scalebox{0.90}{
\begin{tabular}{|c|ccccc|ccc|ccc|}
\hline
Group & LP$_\textsc{Root}$ & GAP$_\textsc{Root}$(\%) & LB & UB & GAP$_\textsc{Final}$(\%) & T(s) & T$_\textsc{Flow}$(s) & T$_\textsc{Root}$(s) & N$_\textsc{Node}$ & N$_\textsc{Tree}$ & S$_\textsc{Tree}$ \\ \hline 
PUC-256-1 & 6559.38 & 7.11 & 6835.63 & 7061.17 & 3.19 & 3600.00 & 241.17 & 34.19 & 86 & - & -\\
PUC-256-2 & 6927.09 & 10.60 & 7423.94 & 7748.26 & 4.19 & 3600.00 & 314.10 & 23.95 & 111 & - & -\\
PUC-256-3 & 6533.74 & 13.10 & 7388.10 & 7518.65 & 1.74 & 3600.00 & 461.95 & 16.17 & 179 & - & -\\
PUC-256-4 & 6724.89 & 10.69 & 7528.98 & 7530.24 & 0.02 & 3600.00 & 410.29 & 14.64 & 199 & - & -\\
PUC-256-5 & 6618.98 & 4.26 & 6913.10 & 6913.38 & $<$0.01 & 3600.00 & 286.83 & 47.27 & 260 & - & -\\
PUC-512-1 & 19460.58 & 17.25 & 19506.06 & 23517.25 & 17.06 & 3600.00 & 354.10 & 380.68 & 7 & - & -\\
PUC-512-2 & 19503.10 & 16.25 & 19549.84 & 23286.79 & 16.05 & 3600.00 & 296.51 & 529.26 & 7 & - & -\\
PUC-512-3 & 18876.79 & 17.48 & 18961.38 & 22874.34 & 17.11 & 3600.00 & 286.42 & 591.55 & 7 & - & -\\
PUC-512-4 & 19305.02 & 17.65 & 19359.15 & 23442.23 & 17.42 & 3600.00 & 187.68 & 1091.63 & 5 & - & -\\
PUC-512-5 & 19510.50 & 14.67 & 19540.73 & 22863.50 & 14.53 & 3600.00 & 189.81 & 1098.31 & 4 & - & -\\
PUC-1024-1 & 56045.21 & 21.08 & 56045.21 & 71011.17 & 21.08 & 3600.00 & - & $>$3600$^\dagger$ & 1 & - & -\\
PUC-1024-2 & 54940.46 & 20.96 & 54940.46 & 69510.15 & 20.96 & 3600.00 & - & $>$3600$^\dagger$ & 1 & - & -\\
PUC-1024-3 & 54861.55 & 20.95 & 54861.55 & 69401.36 & 20.95 & 3600.00 & - & $>$3600$^\dagger$ & 1 & - & -\\
PUC-1024-4 & 53552.76 & 21.29 & 53552.76 & 68040.88 & 21.29 & 3600.00 & - & $>$3600$^\dagger$ & 1 & - & -\\
PUC-1024-5 & 55196.42 & 21.02 & 55196.42 & 69884.32 & 21.02 & 3600.00 & - & $>$3600$^\dagger$ & 1 & - & - \\ \hline
\end{tabular}
}
\vspace*{0.2cm}
\begin{flushleft}
\scriptsize
\ \ $\dagger$ -- The solution of the root node did not terminate
\end{flushleft}
\end{table}
\end{landscape}
}

%%%%%%%%%%%%%%%%%
\end{document}